\documentclass[12pt]{article}
\usepackage{fullpage,epsfig,graphics,amsbsy,amssymb,cancel,slashed,mathrsfs}
\usepackage{psfrag,hyperref}
\usepackage{graphicx,color}
\usepackage{wrapfig}
\usepackage{subfigure}

\newcommand{\be}{\begin{eqnarray}}
\newcommand{\ee}{\end{eqnarray}}
\usepackage{amsmath}
\numberwithin{equation}{section}
\usepackage{caption}
\usepackage[nosort]{cite}
 \usepackage[bulletsep]{collref}

\def\al{\alpha}
\def\eps{\epsilon}

\newcommand{\bea}{\begin{eqnarray}}
\newcommand{\eea}{\end{eqnarray}}  
\newcommand{\nn}{\nonumber}
\newcommand{\Tr}{\textrm{Tr}}

\newcommand{\NN}{\mathcal{N}}

 \newcommand{\sfrac}[2]{\mbox{$\frac{#1}{#2}$}}

\newcommand{\Li}{\mbox{Li}}
\newcommand{\LL}{{\mathcal L}}
\newcommand{\DD}{{\mathcal D}}

\newcommand{\ve}{\varepsilon}
\newcommand{\alggen}[1]{\mathfrak{#1}}
\newcommand{\algS}{\alggen{S}}

\begin{document}

\thispagestyle{empty}
\begin{flushright} \small
UUITP-07/15
 \end{flushright}
\smallskip
\begin{center} \LARGE
{\bf  Gauge theories with 16 supersymmetries on spheres}
 \\[12mm] \normalsize
{\bf  Joseph A. Minahan and Maxim Zabzine} \\[8mm]
 {\small\it
  Department of Physics and Astronomy,
     Uppsala university,\\
     Box 516,
     SE-75120 Uppsala,
     Sweden\\
   }
  
  \medskip 
   \texttt{ joseph.minahan  \& maxim.zabzine  @physics.uu.se}

\end{center}
\vspace{7mm}
\begin{abstract}
 \noindent   We give a unified approach to localization of maximally symmetric gauge theories on spheres, including $S^6$ and $S^7$.  The approach follows Pestun's method of dimensionally reducing from 10 dimensional super Yang-Mills.  The resulting theories have a reduced $R$-symmetry which includes an $SU(1,1)$ subgroup, except in four dimensions where, because of conformal invariance, the full flat-space $R$-symmetry is maintained, and in seven dimensions where $SU(1,1)$ {\it is} the flat-space $R$-symmetry.    For the case of $S^6$ and $S^7$ {we discuss the localization of these theories} 
  and  also present new results for the corresponding matrix models.  The matrix models for $S^6$ and $S^7$ are qualitatively similar to the matrix models of a vector multiplet on $S^4$ and $S^5$ respectively.   We also discuss the contributions of instantons in the six and seven dimensional cases.
 \end{abstract}

\eject
\normalsize

\tableofcontents

 \section{Introduction}
 
Supersymmetric Yang-Mills theories on spheres are useful for investigating aspects of the AdS/CFT correspondence.  Because of the supersymmetry, it is possible to localize the partition function and hence compute certain observables exactly.  These can then be compared to predictions coming from the supergravity dual of the theory.  In the case of $d=4$ $\NN=4$ SYM, Pestun proved \cite{Pestun:2007rz} a conjecture of Erickson, Semenoff, and Zarembo \cite{Erickson:2000af}  and Drukker and Gross \cite{Drukker:2000rr} by showing that the theory localizes to a Gaussian matrix model.  Pestun's results are powerful enough to generalize to $\NN=2$ SYM as well, albeit with a more complicated matrix model.

Generalizations to other dimensions are also possible.  In \cite{Kallen:2012cs,Kallen:2012va} it was shown how to localize $\NN=1$ in five dimensions with a vector multiplet and hypermultiplets in arbitrary representations.  In \cite{Kim:2012ava} it was also shown how to extend this to $\NN=2$, the maximal case for $d=5$.  Going down a dimension, there are an impressive number  of results for three dimensional Chern-Simons theories that are derived using localization, starting with the work in \cite{Kapustin:2009kz} and to  more general theories in \cite{Jafferis:2010un}.

 We can construct supersymmetric theories in other dimensions as well.  The main guidepost is the existence of an acceptable superalgebra, which turns out to be closely related to Nahm's classification of allowed superconformal theories \cite{Nahm:1977tg}.  Basically, if there can be a superconformal theory in $d$ dimensions, then one can have supersymmetry on a sphere in one higher dimension, given the close relation between the $d$-dimensional conformal group $SO(2,d)$ and the $S^{d+1}$ rotation group $SO(d+2)$.  Since $d=6$ is the maximum for a superconformal theory, then $d=7$ is the maximum for a sphere with supersymmetry.

One concern in these constructions is that unless the  theory is also  superconformal  it cannot be supersymmetric without giving up reflection positivity \cite{Festuccia:2011ws} (see also \cite{Kehagias:2012fh}).  Formally, one can get a real supersymmetric Lagrangian by assuming that the $R$-symmetry group is non-compact and the masses of hypermultiplets are imaginary.  In this case, the Euclidean action is unbounded below.  However,  one can analytically continue the scalar field with the wrong sign and end up with a well-defined answer.  This is the approach we take here.

One goal of this paper is to give a more unified approach to localizing gauge theories in different dimensions.  We mainly consider those theories with maximal supersymmetry, which can be obtained by dimensionally reducing ten dimensional super Yang-Mills.  We follow the approach in \cite{Pestun:2007rz} where the dimensionally reduced directions include the time direction.  The gauge fields are Euclidean, but the scalars are Lorentzian.  One advantage of this method is that the fermions retain their real character from the original ten-dimensional theory.  

With this approach we can consider spheres for arbitrary dimension $d$, with $d\le7$.  Localization requires an off-shell formulation, which can  be nicely generalized  to any $d$.  Once the partition function is localized, we can evaluate the action  at the fixed point locus and determine the contributions of the fluctuations.  There is a qualitative difference between the computation of the fluctuation determinant in odd as opposed to even dimensions, namely because the former has a nowhere vanishing vector field while the latter does not.   We will also show how one can reduce to 8 and 4 supersymmetries for low enough dimensions by modifying the actions and off-shell supersymmetry transformations.

We focus on the determinant factors for $d=7$ and $d=6$.  In flat space 16 supersymmetries is the minimal amount for $d=7$ but  for $d=6$  it is possible to have only 8 supersymmetries.  However, it is not clear how to localize without having at least one scalar mode in a vector multiplet, so we are effectively limited to 16 supersymmetries for $d=6$ as well. We will find that the determinant for $d=7$ ($d=6$) is  similar to the determinant for the pure vector theory with 8 supersymmetries in $d=5$ ($d=4$). This further means that the behavior of the eigenvalue distribution as a function of the coupling is also similar to the vector theories in two fewer dimensions. 

In the cases of $d=7$ and $d=6$ we also find the full localization locus, including instanton factors.  In the $d=6$ case  the locus appears to allow for two types of instantons, point-like and extended co-dimension two instantons.  However, we will conjecture that in the end only the point-like instantons  contribute to the partition function.  A similar story holds for the $d=7$ theory.  Here there can be instantons that are  concentrated  along closed $U(1)$ orbits, as well as extended co-dimension three instantons.  We conjecture that only the former contribute to the partition function.

The paper is organized as follows: In section \ref{s-onshell-susy}  we construct the  on-shell supersymmetric transformations and the Lagrangian for maximally supersymmetric Yang-Mills on $S^d$ with $d\le7$.  We also show how to reduce to 8 supersymmetries if $d\le 5$ and 4 supersymmetries if $d\le3$.  In  section \ref{s-offshell-susy} we generalize this construction to an off-shell formalism.   This makes it necessary to introduce seven auxiliary fields and their corresponding pure spinors, with appropriate reductions when the supersymmetry is reduced.  In
section \ref{s-cohomology} we rewrite the supersymmetry transformation in cohomological form.   In section \ref{s-7D}  we specialize to the case of the seven-sphere. We present the detailed calculation 
 of the determinants for $S^7$ with all equivariant parameters turned on. Moreover we conjecture the form of the full answer based on the factorization properties of the perturbative answer. 
In section \ref{s-6D} we present the detailed calculation for $S^6$ and following an analogy with $S^4$ we conjecture the full answer. 
In section \ref{s-matrix} we briefly discuss the properties of 6D and 7D matrix models and argue for $N^2$-behaviour at large $N$.   
In section \ref{s-summary} we summarize our results and discuss some open problems. 

 \section{On-shell supersymmetric gauge theories on spheres.}
 \label{s-onshell-susy}
 
In this section we construct gauge theories which preserve maximal supersymmetry on-shell.  This was first done by Blau in \cite{Blau:2000xg} but we include it here for completeness.  We will closely follow Pestun's construction \cite{Pestun:2007rz} by starting with a 10-dimensional theory and do a Scherk-Schwarz compactification down to a $d$-dimensional sphere.

In 10-dimensional notation, the fields will consist of a gauge field $A_M$, $M=0\dots 9$, and a Majorana-Weyl fermion field, $\Psi_\al$, with $\al=1\dots16$.  The $\Psi_\al$ are assumed to transform in the adjoint representation of the gauge group.  We also use 10-dimensional Dirac matrices ${\Gamma^M}^{\al\beta}$ and $\tilde\Gamma^M_{\alpha\beta}$, where all $\Gamma^M$ and $\tilde\Gamma^M$ are real and symmetric.  More properties and conventions are found in appendix \ref{conv}.  From now on we drop spinor indices.

 The 10-dimensional flat-space Lagrangian is given by \cite{Brink:1976bc}
 \be\label{LL}
\LL= \frac{1}{g_{10}^2}\Tr\left(\sfrac12F_{MN}F^{MN}-\Psi\slashed{D}\Psi\right)\,.
 \ee
 This action is invariant under the supersymmetry transformations
 \be\label{susy}
 \delta_\eps A_M&=&\eps\,\Gamma_M\Psi\,,\nn\\
  \delta_\eps \Psi&=&\sfrac12 \Gamma^{MN}F_{MN}\,\eps\,,
 \ee
 where $\Gamma^{MN}\equiv \tilde\Gamma^{[M}\Gamma^{N]}$.  We will later need the reverse combination, $\tilde\Gamma^{MN}\equiv \Gamma^{[M}\tilde\Gamma^{N]}$. The bosonic supersymmetry parameter $\eps$ is any constant real spinor, hence there are 16 independent supersymmetries.
 
 We next dimensionally reduce to a $d$-dimensional Euclidean gauge theory.  Hence, we will have gauge fields, $A_\mu$, $\mu=1\dots d$, as well as scalar fields $\phi_I\equiv A_I$, where we choose $I=0, d+1,\dots 9$.  We assume that all derivatives along  compactified directions are zero, so that the field-strengths become,
 \be
 F_{\mu I}&=&[D_\mu,\phi_I]\nn\\
 F_{IJ}&=&[\phi_I,\phi_J]\,.
 \ee
 Since the $0$-direction is time-like, $\phi_0$ will have the wrong sign for its kinetic term.  Assuming that the $d$-dimensional space is flat Euclidean, the scalars transform under the vector representation of an $SO(1,9-d)$ $R$-symmetry group.  We can also decompose the fermions $\Psi$ under the reduction, but it will not be necessary for our purposes here, at least while maintaining maximal supersymmetry.  The coupling in the dimensionally reduced theory is $g_{YM}^2=g_{10}^2/V_{10-d}$, where $V_{10-d}$ is the volume of the compactified space.
 
 We next assume that instead of flat-space, our $d$-dimensional space is the sphere $S^d$ with radius $r$.  In 4 dimensions the gauge theory is superconformal, which means that we must add to the action the conformal mass term
 \be
S_{\phi\phi}= \frac{1}{g_{YM}^2}\int d^4x\sqrt{-g}\left(\frac{2}{r^2}\,\Tr \phi_I\phi^I\right)~.
 \ee
In other dimensions the theory is not conformal, but we  still include a similar term for the scalar fields
 \be\label{Scc}
S_{\phi\phi}= \frac{1}{g_{YM}^2}\int d^dx\sqrt{-g}\left(\frac{d\,\Delta_I}{2\,r^2}\,\Tr \phi_I\phi^I\right)\,,
 \ee
 where the index $I$ is summed over and $\Delta_I$ is the analog of the dimension for $\phi_I$.
We will see that we need further terms to preserve the supersymmetry. 

On the sphere we can no longer have supersymmetries defined by  constant spinors.  To find the desired spinors we instead consider conformal Killing spinors   (CKS) that satisfy the equations
\be\label{CKSE}
\nabla_\mu\eps=\tilde\Gamma_\mu \tilde\eps\,,\qquad \nabla_\mu\tilde\eps=-\frac{1}{4r^2}\Gamma_\mu\eps\,.
\ee
Note that $\Gamma_\mu=e_{\hat\mu\mu}\Gamma^{\hat\mu}$, where $\Gamma^{\hat\mu}$ is a Minkowskian 10-dimensional $\Gamma$-matrix along the sphere directions.  Following \cite{Pestun:2007rz}, and writing the sphere metric as
\be
ds^2=\frac{1}{(1+\frac{x^2}{4r^2})^2}\, dx^2\,,
\ee
a general solution for (\ref{CKSE}) is
\be\label{GKS}
\eps=\frac{1}{(1+\frac{x^2}{4r^2})^{1/2}}\left(\eps_s+x\cdot \Gamma\tilde\eps_c\right)\,,
\ee
where $\eps_s$ and $\tilde\eps_c$ are arbitrary constant spinors.  Hence, there are 32 independent CKS's.  We can reduce these to 16 spinors by also imposing,
\be\label{KS}
\nabla_\mu\eps=\beta\tilde\Gamma_\mu\Lambda \eps\,,
\ee
where $\beta=\frac{1}{2r}$.  Consistency with (\ref{CKSE}) requires that $\Lambda\tilde\Gamma^\mu\Lambda=-\Gamma^\mu$.  Shortly we will  find that  in order to build a supersymmetric Lagrangian for $d\ne4$ we must set $\Lambda^T=-\Lambda$.  The simplest choice that satisfies these conditions has $\Lambda= \Gamma^8\tilde\Gamma^9\Gamma^0$, allowing  this construction  for  spheres up to dimension 7.  With this additional condition, the constant spinors in (\ref{GKS}) satisfy $\tilde\eps_c=\beta\Lambda\eps_s$, and thus reduces the number of independent   spinors to 16.

The supersymmetry transformations in (\ref{susy}) need to be modified.  To this end, we propose the ansatz
 \be\label{susysp}
 \delta_\eps A_M&=&\eps\,\Gamma_M\Psi\,,\nn\\
  \delta_\eps \Psi&=&\sfrac12 \Gamma^{MN}F_{MN}\eps+\frac{\alpha_I}{2}\Gamma^{\mu I}\phi_I\nabla_\mu\,\eps\,,
 \ee
 where the index $I$ is again summed over and the constants $\alpha_I$ are given by
 \be\label{alrel}
\alpha_I&=&\frac{4(d-3)}{d}\,,\qquad I=8,9,0\nn\\
\alpha_I&=&\frac{4}{d}\,,\qquad I=d+1,\dots 7\, .
\ee
To distinguish the different types of internal indices, we will use $\phi_A$ and $\phi_i$, with $A=8,9,0$, and $i=d+1,\dots 7$.  For $d=4$, we see that this is precisely the modification in \cite{Pestun:2007rz}.  With these modifications the change to the Lagrangian, including the extra term from (\ref{Scc}), is
\be\label{dL}
g_{YM}^2\,\delta\LL&=&\Tr\bigg[-(d-4)F^{\mu\nu}\tilde\eps\Gamma_{\mu\nu}\Psi-(d(2-\alpha_I)-4)F^{I\nu}\tilde\eps\Gamma_{I\nu}\Psi\nn\\
&&\qquad-d\left(1-\frac{\alpha_I+\al_J}{2}\right)F^{IJ}\tilde\eps\Gamma_{IJ}\Psi-\frac{d}{r^2}\left(\frac{d\al_I}{4}-\Delta_I\right)\phi_I\eps\Gamma^I\Psi\bigg]\,,
\ee
up to total derivatives. 
If $d=4$ then $\alpha_I=\Delta_I=1$ for all $I$ and the Lagrangian is invariant.  For $d\ne4$ we need further modifications to $\LL$.  

We start by adding the term
\be\label{Lpsi}
\LL_{\Psi\Psi}=\frac{1}{g_{YM}^2}(d-4)\beta \Tr\Psi\Lambda\Psi\,,
\ee
which is nontrivial only if $\Lambda^T=-\Lambda$.
Under a supersymmetry transformation (\ref{Lpsi}) changes by
\be\label{dLpsi}
g_{YM}^2\delta\LL_{\Psi\Psi}&=&\Tr\bigg[-(d-4)F^{\mu\nu}\beta\eps\tilde\Gamma_{\mu\nu}\Lambda\Psi-2(d-4)\beta F^{I\nu}\eps\tilde\Gamma_{I\nu }\Lambda\Psi\nn\\
&&\qquad\qquad-(d-4)\beta F^{IJ}\eps\tilde\Gamma_{IJ}\Lambda\Psi+(d-4)\beta^2\al_I d\, \phi_I \eps \Lambda \tilde \Gamma^I\Lambda \Psi\bigg]\,.
\ee
Using (\ref{KS}) we see that the first term in (\ref{dLpsi}) cancels the first term in (\ref{dL}).  Using 
$\Lambda\Gamma_{A\nu}=-\tilde\Gamma_{A\nu}\Lambda$ and $\Lambda\Gamma_{i\nu}=+\tilde\Gamma_{i\nu}\Lambda$ we see that the second terms cancel given the $\al_I$  in (\ref{alrel}).  

The relations $\Lambda\Gamma_{ij}=+\tilde\Gamma_{ij}\Lambda$ and $\Lambda\Gamma_{iA}=-\tilde\Gamma_{iA}\Lambda$ lead to the cancelation of the third terms for the $ij$ and $iA$ combinations.  However, the $AB$ combination has a leftover piece
\be
-4(d-4)\beta \Tr\left[F^{AB}\eps\tilde\Gamma_{AB}\Lambda\Psi\right]=4(d-4)\beta \Tr\left[[\phi^A,\phi^B]\eps \Gamma^C\Psi\right]\varepsilon_{ABC}\,,
\ee
where $\varepsilon_{890}=1$.  This term can be canceled by including the extra term in the Lagrangian
\be\label{LABC}
\LL_{ABC}=-\frac{1}{g_{YM}^2}\frac{2}{3r}(d-4)\Tr([\phi^A,\phi^B]\phi^C)\varepsilon_{ABC}\,.
\ee

Finally, the fourth terms from (\ref{dLpsi}) and (\ref{dL}) combine to give
\be
\frac{d}{r^2}\Tr\left[(\Delta_A-\alpha_A)\phi_A\eps\Gamma^A\Psi+(\Delta_i-(d/2-1)\al_i)\phi_i\eps\Gamma^i\Psi\right]\,,
\ee
This is zero if $\Delta_A=\al_A$ and $\Delta_i=2(d-2)/d$.
Hence, the complete supersymmetric Lagrangian is 
\be\label{Lss}
\LL_{ss}&=&\frac{1}{g_{YM}^2}\Tr\Bigg(\sfrac12F_{MN}F^{MN}-\Psi\slashed{D}\Psi+\frac{(d-4)}{2r}\Psi\Lambda\Psi+\frac{2(d-3)}{r^2} \phi^A\phi_A+\frac{(d-2)}{r^2}\phi^i\phi_i\nn\\
&&\qquad\qquad\qquad\qquad -\frac{2}{3r}(d-4)[\phi^A,\phi^B]\phi^C\varepsilon_{ABC}\Bigg)\,.
\ee
Notice that for $d\ne4,7$  the $R$-symmetry is broken from $SO(1,9-d)$ to a smaller group that contains an $SU(1,1)$ subgroup.  In the case of $d=5$, (\ref{Lss}) is equivalent to the $\NN=2$ Lagrangian in \cite{Kim:2012ava}, after a Euclidean rotation of $\phi^0$. 

 If $d\le5$, we can split up $\Psi$ into parts that are eigenstates of $\Gamma\equiv\Gamma^{6789}$, with the even eigenstates, $\psi=+\Gamma\psi$ making up the fermions in the vector multiplet and the odd eigenstates, $\chi=-\Gamma\chi$ making up those in an adjoint hypermultiplet.  The scalars $\phi^I$,  $I=6\dots 9$ are the bosonic fields in the hypermultiplet. The gauge fields $A_\mu$ and the other scalar fields, $\phi_I$, $I=0,d+1, \dots 5$ are the bosonic fields in the vector multiplet.  By also assuming that $\eps$ is chiral, $\eps=+\Gamma\eps$, we reduce the number of supersymmetries to 8.   Note that this  reduction does not work for $d=6$, even though it is possible to have 8 supersymmetries, at least in flat-space.  In this case one would have to use $\Gamma^{0789}$, but its square is $-1$, and since $\Psi$ is assumed real, it cannot split into eigenstates of $\Gamma^{0789}$.
 
The breaking of the $R$-symmetry to $SU(1,1)$  can also be understood from the available superalgebras.  In the case of $S^6$  the superalgebra is the exceptional  $F_4$ which has 16 supercharges and an $SO(7)\times SU(1,1)$ bosonic subalgebra.  This can be compared to the 5-dimensional superconformal algebra which has the bosonic subgroup $SO(2,5)\times SU(2)$.  Similarly, for $S^7$ we have the superalgebra $OSp(8|2,\mathbb{R})$ which also has 16 supercharges and is closely related to the $6$-dimensional $(1,0)$ superconformal algebra $OSp(2,6|2)$.

 With this reduction we can relax the conditions on the additional terms that are added to the Lagrangian.  If we replace $\Psi$ with $\psi$ in (\ref{dL}) and (\ref{dLpsi}), there are only contributions from terms with an even number of hypermultiplet scalar indices.  This leaves the values of $\al_I$ and $\Delta_I$ unchanged for the vector multiplet indices.  In fact, the only place the hypermultiplet scalars enter into these transformations is in the third terms, where the constants come in pairs $\al_I+\al_J$ in (\ref{dL}).  This suggests modifying the transformations such that $\al_I+\al_J$ stays fixed, at least for some combination of the $I$ and $J$.  To this end, we let
\be\label{al8}
\al_I&=&\frac{2(d-2)}{d}+\frac{4i\sigma_I \,m\,r}{d}\qquad I=6\dots 9\nn\\
\sigma_I&=&+1\qquad\qquad I=6,7\nn\\
\sigma_I&=&-1\qquad\qquad I=8,9\,,
\ee
where $m$ serves as the hypermultiplet mass.  In the case that $I,J$ has one index from the pair $6,7$ and one index from the pair $8,9$, the linear combination $\al_I+\al_J$ is unchanged from the maximally supersymmetric case and the third terms cancel. If $I,J=6,7$ or $I,J=8,9$, then
\be
-d\left(1-\frac{\alpha_I+\al_J}{2}\right)\eps\tilde\Gamma_{IJ}\Lambda\psi=\sigma_I(d-4+4i\sigma_I m\,r)\eps\Gamma^0\psi\,.
\ee
The corresponding term can be canceled by adding to the Lagrangian 
\be\label{3phi}
\LL_{\phi\phi\phi}=\frac{1}{g_{YM}^2}\left(\left(\frac{2(d-4)}{r}+4im\right)\Tr(\phi^0[\phi^6,\phi^7])-\left(\frac{2(d-4)}{r}-4im\right)\Tr(\phi^0[\phi^8,\phi^9])\right)\,.
\nn\\
\ee
In 5 dimensions we see that if $mr=i/2$ the first term goes away and the second term matches (\ref{LABC}).

If we now replace $\Psi$ with $\chi$, we see that the first terms in (\ref{dL}) and (\ref{dLpsi}) do not contribute, hence the restriction on the coefficient of the $\chi\Lambda\chi$ term is removed.  The only terms that contribute from the rest of (\ref{dL}) and (\ref{dLpsi}) contain exactly one hypermultiplet scalar. It is easy to check that with the assignment in (\ref{al8}), the Lagrangian should contain the hypermultiplet mass term
\be\label{Lchi}
\LL_{\chi\chi}=\frac{1}{g_{YM}^2}\left(- im\Tr \chi\Lambda\chi\right)\,,
\ee
in order for the second and third terms to cancel (with the coefficient $\beta(d-4)$ in (\ref{dLpsi}) replaced with $-im$).  In order to cancel the fourth terms we must choose
\be\label{Delta8}
\Delta_I=\frac{2}{d}\left(mr(mr+i\sigma_I)+\frac{d(d-2)}{4}\right)\,.
\ee
Hence, the complete Lagrangian with an adjoint hypermultiplet is the dimensionally reduced version of (\ref{LL}) and supplemented with
\be
\LL_{\phi\phi}+\LL_{\phi\phi\phi}+\LL_{\psi\psi}+\LL_{\chi\chi}\,,
\ee
where the terms are defined in (\ref{Scc}), (\ref{3phi}), (\ref{Lpsi}) and (\ref{Lchi}), and where $\LL_{\psi\psi}$ replaces $\Psi$ with $\psi$ in (\ref{Lpsi}).   This is of course equivalent to the construction in \cite{Hosomichi:2012ek}.

If $d\le3$ we can follow a similar procedure to reduce the number of supersymmetries to 4.  We  define a second matrix $\Gamma'=\Gamma^{4589}$, in which case $\Psi$ can split into $\Gamma'\psi=\Gamma\psi=+\psi$, and $\chi_{(j)}$, $j=1,2,3$ where $\Gamma\chi_{(j)}=(-1)^{\beta_2(j)}\chi_{(j)}$, $\Gamma'\chi_{(j)}=(-1)^{\beta_1(j)}\chi_{(j)}$,  and where $\beta_k(j)$  are the binary digits for $j$, {\it i.e.}  $j= 2\cdot\beta_2(j)+\beta_1(j)$.  The $\chi_{(j)}$ are the fermionic part of chiral multiplets while the $\psi$ is the fermionic part of the vector multiplet.  The supersymmetries are reduced to 4 by also setting $\Gamma\eps=\Gamma'\eps=\eps$.  With these choices we can then set
\be\label{al9}
\al_{(j)}&=&\frac{2(d-2)}{d}+\frac{4i\sigma_{(j)} \,m_j\,r}{d}\,,\qquad
\sigma_{(j)}=(-1)^{\beta_2(j)\beta_1(j)}\nn\qquad\\
&&\al_{4}=\al_5=\al_{(1)}\,,\quad\al_6=\al_7=\al_{(2)}\,,\quad\al_8=\al_9=\al_{(3)}\,.
\ee
The other $\al_I$ keep their same assignments.  The $\Delta_I$ change in the same way as in (\ref{Delta8}) and become
\be\label{Delta4}
\Delta_I=\frac{2}{d}\left(m_jr(m_jr+i\sigma_{(j)})+\frac{d(d-2)}{4}\right)\,,
\ee
while the mass terms for the  chiral multiplet fermions are
\be\label{Lchi4}
\LL_{\chi\chi}=\frac{1}{g_{YM}^2}\left(- im_j\Tr \chi_{(j)}\Lambda\chi_{(j)}\right)\,.
\ee
Freed from the vector multiplet we can add more chiral multiplets, or change their representations, {\it etc.}

  \section{Off-shell supersymmetry and localization}
  \label{s-offshell-susy}

 In order to localize we need an off-shell formulation of the supersymmetry transformations.  A Euclidean version of this was done for the maximally supersymmetric case in \cite{Fujitsuka:2012wg}.  An off-shell formulation does not exist for  all 16 supersymmetries, but it is only necessary to go off-shell for one particular $\epsilon$.  The choice of $\epsilon$ is specified by how one chooses the vector field $v^M$,
 \be\label{veq}
 v^M\equiv \eps\Gamma^M\eps\,.
 \ee
 In the appendix it is shown that $v^Mv_M=0$.  
   Again following \cite{Pestun:2007rz}, for a choice of $\epsilon$ we  also choose seven bosonic pure-spinors $\nu_m$, $m=1\dots 7$, and an accompanying set of auxiliary fields $K^m$.    The pure-spinors satisfy the orthogonality relations
 \be\label{psprop}
 \eps\Gamma^M\nu_m&=&0\nn\\
 \nu_m\Gamma^M\nu_n&=& \delta_{nm}v^M\,.
 \ee

The off-shell supersymmetry transformations are then modified to
 \be\label{susyos}
 \delta_\eps A_M&=&\eps\,\Gamma_M\Psi\,,\nn\\
  \delta_\eps \Psi&=&\sfrac12 \Gamma^{MN}F_{MN}\eps+\frac{\alpha_I}{2}\Gamma^{\mu I}\phi_I\nabla_\mu\,\eps+K^m\nu_m\, ,\nn\\
\delta_\eps K^m&=&-\nu^m\slashed{D}\Psi+\Delta K^m\,,
 \ee
 where the coefficients $\al_I$ have the same values as in the previous section.  The transformation term $\Delta K^m$ is an extra piece that is required if $d\ne4$.
 
 It is necessary to show that the transformations in (\ref{susyos}) close up to a symmetry of the action in order for the supersymmetry to hold.  Starting with the $A_\mu$ fields, we have that
 \be
 \delta^2_\eps A_\mu&=&\delta_\eps(\eps\Gamma_\mu \Psi)=\frac12F^{MN}\eps\Gamma_\mu \Gamma_{MN}\eps-\frac{1}{2}\beta d \al_I\phi_I\eps\Gamma_\mu\tilde\Gamma^I\Lambda\eps+K^m\eps\Gamma_\mu\nu_m\, .
 \ee
The last term in the rhs is zero by (\ref{psprop}) while the second is zero if $I$ is one of the $A$ indices, since two $\eps$ spinors sandwiching three distinct $\Gamma^M$ matrices is zero by antisymmetry.  Hence, after contracting some $\Gamma^M$ matrices we can write
\be
  \delta^2_\eps A_\mu&=&\eps\Gamma^N\eps F_{\mu N}+2\phi_I\eps\Gamma^I\nabla_\mu\eps\nn
 \\
 &=&-v^\nu F_{\nu\mu}+[D_\mu,v^I\phi_I]\,.
 \ee
 The first term is the negative of the Lie derivative along the vector field $v$, while the second term is a gauge transformation, hence this closes up to symmetries of the action.
 
 A similar calculation shows that
 \be\label{phit}
 \delta^2_\eps\phi_I=-v^\nu D_\nu\phi_I-[v^J\phi_J,\phi_I]-\frac{1}{2}\al_I\beta d\, \eps\tilde\Gamma_{IJ}\Lambda\eps\,\phi^J\,,
 \ee
 where the first two terms are again the negative of the Lie derivative and a gauge transformation.  The last term is an $R$-symmetry transformation.  Notice that this term is non-zero only if $I$ and $J$ are both $A$ type indices or both $i$ type indices.  Hence, this is the unbroken part of the $R$-symmetry on the sphere, and so this too is a symmetry of the action. 
 
 The closure of $\Psi$ is more involved.  Using the triality identity in (\ref{triality}) and the last relation in (\ref{psrel}), we can reduce the transformation to
 \be
 \delta^2_\eps\Psi&=&-v^N D_N\Psi-\beta(\eps\Lambda\tilde\Gamma_\mu\Gamma_N\Psi)\Gamma_{\mu N}\eps-2(d-3)\beta(\eps\Gamma_A\Psi)\tilde\Gamma^A\Lambda\eps-2\beta(\eps\Gamma_i\Psi)\tilde\Gamma^i\Lambda\eps\nn\\
 &&\qquad\qquad\qquad\qquad\qquad\qquad+\Delta K^m\nu_m\,.
 \ee
Using the triality identity again, we can further manipulate this to
 \be\label{psit}
 \delta^2_\eps\Psi&=&-v^N D_N\Psi-\frac{1}{4}(\nabla_{[\mu}v_{\nu]})\Gamma^{\mu\nu}\Psi-\frac{1}{2}\beta(\eps\Lambda\Gamma^{IJ}\eps)\Gamma_{IJ}\Psi\nn\\
 &&\qquad+(d-4)\beta\left((\eps\Lambda\Psi)\eps+(\eps\Gamma_N\Psi)\tilde\Gamma^N\Lambda\eps-2\,(\eps\Gamma_A\Psi)\tilde\Gamma^A\Lambda\eps\right)+\Delta K^m\nu_m\, .
 \ee
 The first two terms combine to give the Lie derivative of a spinor, which includes a spin-connection, and the gauge transformation. 
 We also have that
 \be
 (\eps\Gamma_N\Psi)\tilde\Gamma^N\Lambda\eps-2\,(\eps\Gamma_A\Psi)\tilde\Gamma^A\Lambda\eps=- (\eps\Gamma_N\Psi)\tilde\Lambda\Gamma^N\eps=\frac12 v^N\tilde\Lambda\Gamma_N\Psi \, .
 \ee
 The form of the first term in the second line of (\ref{psit}) suggests that one should set
 \be\label{dKm}
 \Delta K^m=\beta(d-4) \nu^m\Lambda\Psi\,,
 \ee
 in order to  put (\ref{psit}) into a nicer form by using  (\ref{psrel}).    Indeed with (\ref{dKm}) for $\Delta K^m$, we find
\be
\delta^2_\eps\Psi&=&-v^N D_N\Psi-\frac{1}{4}(\nabla_{[\mu}v_{\nu]})\Gamma^{\mu\nu}\Psi-\frac{1}{2}\beta(\eps\Lambda\Gamma^{IJ}\eps)\Gamma_{IJ}\Psi\nn\\
 &&\qquad\qquad\qquad+(d-4)\beta(\eps\Gamma^A\eps)\tilde\Gamma_A\Lambda\Psi\nn\\
 &=&-v^N D_N\Psi-\frac{1}{4}(\nabla_{[\mu}v_{\nu]})\Gamma^{\mu\nu}\Psi\nn\\
&&\qquad\qquad -\frac{1}{2}(d-3)\beta(\eps\tilde\Gamma^{AB}\Lambda\eps)\Gamma_{AB}\Psi- \frac{1}{2}\beta(\eps\tilde\Gamma^{ij}\Lambda\eps)\Gamma_{ij}\Psi\,.
  \ee
  The last term is consistent with the $R$-symmetry transformation in (\ref{phit}).  Note that with (\ref{dKm}) the transformation for $K^m$ in (\ref{susyos}) is proportional to the modified equation of motion for $\Psi$.
  
  Finally, the closure for transformations on $K^m$ can be confirmed using (\ref{dKm}), where one finds after similar manipulations of the expressions,
  \be
 \delta^2_\eps K^m=-v^MD_M K^m-(\nu^{[m}\Gamma^\mu\nabla_\mu \nu^{n]})K_n+(d-4)\beta (\nu^{[m}\Lambda\nu^{n]})K_n\,.
 \ee
 The last two terms combine to generate internal $SO(7)$ transformations amongst the $K^m$.  Hence, the  off-shell action should be invariant under these  transformations.
 
 We now look for the Lagrangian invariant under these transformations.   In 4-dimensions one modifies the Lagrangian by adding the auxiliary field piece \cite{Pestun:2007rz}
 \be
 \LL_{aux}=-\frac{1}{g_{YM}^2}\,\Tr K^mK_m\,,
 \ee
 which is  invariant under the internal $SO(7)$ symmetry.  We take this as an ansatz for all $d$ and add it to the Lagrangian in (\ref{Lss}).  The only terms we need to check in the transformation are those that are linear in $K^m$.  To this end, we need to verify that
 \be
 \delta_\eps^K\left(-\Psi\slashed{D}\Psi+(d-4)\beta\Psi\Lambda\Psi\right)-\delta_\eps(K^mK_m)=0\,.
 \ee
 This is clearly true based on the transformations in (\ref{susyos}) and (\ref{dKm}), hence the Lagrangian $\LL_{ss}+\LL_{aux}$ is invariant.
 
We now turn to localizing the off-shell action. We follow the usual procedure of modifying the path integral to
\be
Z=\int\DD\Phi e^{-S-t Q V}\,,
\ee
where $Q$ is a fermionic symmetry generator, and  $\DD\Phi$ represents the functional integral measure for all fields.  Since $Q$ is assumed to be a symmetry of the action and the measure, an elementary argument shows that the path integral is independent of $t$.  One then takes $t\to\infty$ so that the  fields in the path integral localize onto the zeros of $QV$.  

For $Q$ we choose the supersymmetry generated by $\eps$, $\delta_\eps$, while for $V$ we use
\be
V=\int d^dx\sqrt{-g} \,\Psi\, \overline{\delta_\eps\Psi}\,.
\ee
$ \overline{\delta_\eps\Psi}$ is the conjugate of the transformation in (\ref{susyos}),
\be
 \overline{\delta_\eps\Psi}=\sfrac12\tilde \Gamma^{MN}F_{MN}\Gamma^0\eps+\frac{\alpha_I}{2}\tilde\Gamma^{\mu I}\phi_I\Gamma^0\nabla_\mu\,\eps-K^m\Gamma^0\nu_m\,.
 \ee
Therefore, the bosonic part of $\delta_\eps V$ is given by
\be
\delta_\eps V\Big|_{bos}=\int d^dx\sqrt{-g} \,\Tr(\delta_\eps\Psi\, \overline{\delta_\eps\Psi})\,.
\ee
Carrying out the expansion of the integrand, we find that many terms are zero.  Those terms that are left-over can be written as
\be\label{bosdPdP}
\delta_\eps\Psi\, \overline{\delta_\eps\Psi}&=&\frac{1}{2}F_{MN}F^{MN}-\frac{1}{4}F_{MN}F_{M'N'}(\eps\Gamma^{MNM'N'0}\eps)\nn\\
&&\qquad+\frac{\beta d\al_I}{4}F_{MN}\phi_I(\eps\Lambda(\tilde\Gamma^I\tilde\Gamma^{MN}\Gamma^0-\tilde\Gamma^0\Gamma^I\Gamma^{MN})\eps)\nn\\
&&\qquad- K^mK_m v^0 -\beta d\, \al_0\phi_0 K^m(\nu_m\Lambda\eps)+\frac{\beta^2d^2}{4}\sum_I(\al_I)^2\phi_I\phi^I v^0\,.
\ee
At this point we 
 assume the consistent choice $v^0=1$, $v^{8,9}=0$, in which case the last line in (\ref{bosdPdP}) can be written as
\be\label{fpeq}
-(K^m+2\beta(d-3)\phi_0(\nu_m\Lambda\eps))^2+\frac{\beta^2d^2}{4}\sum_{J\ne 0}(\al_J)^2\phi_J\phi^J\,,
\ee
where we used (\ref{psrel}) to write the first term as a square.
(\ref{fpeq}) then leads to the fixed-point locus
\be\label{fpl}
K^m=-2\beta(d-3)\phi_0(\nu_m\Lambda\eps)\,,\qquad \phi_J=0\ \ {J\ne0}\,.
\ee
The first term in (\ref{bosdPdP}) contains the kinetic terms for $\phi_0$, therefore the localization restricts $\phi_0$ to be covariantly constant on the sphere.  The rest of the fixed point conditions allow for
 generalized instantons, but we will return to this point later when we specialize to the various dimensions.

If we now substitute the values at the fixed point into the Lagrangian $\LL$, we find, assuming the gauge fields are zero,
\be\label{LLfp}
\LL_{fp}=-\frac{1}{g_{YM}^2}\frac{(d-1)(d-3)}{r^2}\Tr(\phi_0\phi_0)\,.
\ee
Hence, the action is 
\be
S_{fp}=V_d\LL_{fp}=-\frac{4\pi^2r^{d-4}S_{d-4}}{g_{YM}^2}\Tr\sigma^2\,,
\ee
where $V_d$ is the volume of the sphere, $\sigma$ is the dimensionless variable $\sigma=r\phi_0$ and $S_{d-4}$ is the volume of a unit $d-4$ sphere.  Note that action is 0 for $d=3$, which is why one needs a Chern-Simons term to get interesting behavior from localization in three dimensions.   Also note that the action is unbounded from below if $d>3$ (or $d<1$),  and therefore we must analytically continue $\sigma$ to the imaginary axis when doing the path integral for $d$ in this range.

To study (\ref{bosdPdP}) further, 
 we note by either antisymmetry or the conditions on $v^M$, that in order for the second or third terms to be nonzero none of their indices  can be $0$.  To this end we assume that the $I$ and $A$ indices are not 0 and write the second term as
 \be\label{FwF2}
&& -\frac{1}{4}F_{MN}F_{M'N'}(\eps\Gamma^{MNM'N'0}\eps)=\nn\\
&&\qquad-\frac{1}{4}F_{\mu\nu}F_{\lambda\rho}(\eps\Gamma^{\mu\nu\lambda\rho0}\eps)-D_\mu\phi_I F_{\lambda\rho}(\eps\Gamma^{\mu I\lambda\rho0}\eps)+\left(D_\mu\phi_ID_\nu\phi_{J}-\frac{1}{2}F_{\mu\nu}[\phi_I,\phi_J]\right)(\eps\Gamma^{\mu\nu IJ0}\eps)\nn\\
&&\qquad\qquad\qquad-D_\mu\phi_I[\phi_J,\phi_K](\eps\Gamma^{\mu IJK0}\eps)-\frac{1}{4}[\phi_I,\phi_J][\phi_K,\phi_L](\eps\Gamma^{ IJKL0}\eps)\\
\label{FwF}&&\qquad=-\frac{1}{4}F_{\mu\nu}F_{\lambda\rho}(\eps\Gamma^{\mu\nu\lambda\rho0}\eps)-aD_\mu\phi_I F_{\lambda\rho}(\eps\Gamma^{\mu I\lambda\rho0}\eps)-
2\beta(1-a)(d-2)\phi_I F_{\lambda\rho}(\eps\Gamma^{ I\lambda\rho89}\eps)\nn\\
&&\qquad\qquad\qquad+2\beta(d-1)\phi_ID_\nu\phi_{J}(\eps\Gamma^{\nu IJ89}\eps)\nn\\
&&\qquad\qquad\qquad-\frac{2\beta d}{3}\phi_I[\phi_J,\phi_K](\eps\Gamma^{ IJK89}\eps)-\frac{1}{4}[\phi_I,\phi_J][\phi_K,\phi_L](\eps\Gamma^{ IJKL0}\eps)\,,
 \ee
 where we integrated by parts and used the Bianchi identity.  For the second term on the right hand side of   (\ref{FwF2}) we split it into two pieces with coefficients $a$ and $1-a$, and only integrate by parts the second piece.  This will be useful when we consider the localization locus on $S^6$.  We next write 
 the third term in (\ref{bosdPdP}) as
 \be\label{DphiF}
&& \frac{\beta d\al_I}{4}F_{MN}\phi_I(\eps\Lambda(\tilde\Gamma^I\tilde\Gamma^{MN}\Gamma^0-\tilde\Gamma^0\Gamma^I\Gamma^{MN})\eps)\nn\\
&&\qquad= 2\beta\Bigg(\phi_i[\phi_j,\phi_k](\eps\Gamma^{ijk89}\eps)-(2d-5)\phi_i[\phi_A,\phi_B]\varepsilon^{AB}v^i+2(d-3)\phi_A D_\mu\phi_B\varepsilon^{AB}v^\mu\nn\\
&&\qquad\qquad\qquad\qquad-2\phi_i D_\mu\phi_j(\eps \Gamma^{\mu ij89}\eps)+\phi_iF_{\mu\nu}(\eps\Gamma^{i\mu\nu89}\eps)\Bigg)\,.
 \ee
 Combining (\ref{FwF}) and (\ref{DphiF}) together, we arrive at
 \be\label{FwFphiF|}
&& -\frac{1}{4}F_{MN}F_{M'N'}(\eps\Gamma^{MNM'N'0}\eps)+ \frac{\beta d\al_I}{4}F_{MN}\phi_I(\eps\Lambda(\tilde\Gamma^I\tilde\Gamma^{MN}\Gamma^0-\tilde\Gamma^0\Gamma^I\Gamma^{MN})\eps)=\nn\\
&&\qquad-\frac{1}{4}F_{\mu\nu}F_{\lambda\rho}(\eps\Gamma^{\mu\nu\lambda\rho0}\eps)-\frac{1}{4}[\phi_I,\phi_J][\phi_K,\phi_L](\eps\Gamma^{ IJKL0}\eps)\nn\\
&&\qquad+2\beta(d-5)\Big(\phi_A D_\mu\phi_B\varepsilon^{AB}v^\mu-\phi_i[\phi_A,\phi_B]\varepsilon^{AB}v^i
\Big)\nn\\
&&\qquad+2\beta(d-3)\Big(\phi_i D_\mu\phi_j(\eps \Gamma^{\mu ij89}\eps)-\frac{1}{3}\phi_i[\phi_j,\phi_k](\eps\Gamma^{ijk89}\eps)\Big)\nn\\
&&\qquad-aD_\mu\phi_I F_{\lambda\rho}(\eps\Gamma^{\mu I\lambda\rho0}\eps)-2\beta(d-3-a(d-2))\phi_i F_{\mu\nu}(\eps\Gamma^{i\mu\nu89}\eps)
\ee
 We will further simplify these expressions in the next sections after specializing to specific dimensions.
 
The off-shell construction for theories with reduced symmetry can be modified in a straightforward manner.  Without going into detail, for 4 supersymmetries we can split the auxiliary fields and the pure spinors  into sets $\kappa_0,\nu_0$ and $\kappa_{(j)}^{s}$, $\nu_{(j),s}$, with $s=1,2$ and $j=1,2,3$.  The pure spinor $\nu_0$ has the same chirality as $\eps$ while  $\nu_{(j),s}$ has the same chirality as $\chi_{(i)}$.  The transformation in the last line of (\ref{susyos}) is then modified to 
 \be\label{susyosk4}
\delta_\eps K_0&=&-\nu_0\slashed{D}\psi+\beta(d-4)\nu_0\Lambda\psi\nn\\
\delta_\eps K_{(j)}^s&=&-\nu_{(j)}^s\slashed{D}\chi_{(j)}-im_j\nu_{(j)}^s\Lambda\chi_{(j)}\,.
 \ee
 The transformations for the fermions stays the same as in (\ref{susyos}), taking into account the modifications to the $\al_I$ and matching the different chiralities to each other.  If we set $m_2=m_3$ and $m_1=i\beta(d-4)$ then we have 8 supersymmetries and the off-shell transformations in $d=5$ are equivalent to \cite{Hosomichi:2012ek}.

 \section{Cohomological complex}
 \label{s-cohomology}

 The spinors $\eps$ in (\ref{KS}) can be used to build other quantities which will be of great use to us.  We have already mentioned the vector vield $v^M\equiv \eps \Gamma^M\eps$, which can be shown to satisfy $v^Mv_M=0$ (see appendix for details about deriving this and other properties.).    As in the previous section, we  choose an $\eps$ such that  $v^0=1$.  In what follows, we need to distinguish between odd and even dimensional spheres.  For odd dimensions it is also possible to choose $v^I=0$, $I\ne 0$.  It then follows that $v^\mu v_\mu=1$, hence there exists an everywhere nonzero vector field on the sphere.  We can identify $v_\mu$ as a contact 1-form $\kappa_\mu$, which we can then use to construct a two-form,
\be\label{kappa-omega}
\nabla_{[\mu}\kappa_{\nu]}=-\frac{1}{r}\eps\tilde\Gamma_{\mu\nu}\Lambda\eps\equiv-\frac{1}{r}\omega_{\mu\nu}\,,
\ee
where we used (\ref{KS}) and the antisymmetry of $\Lambda$.  It is also straightforward to show that $v^\mu\omega_{\mu\nu}=0$ and $\nabla_{[\mu}\omega_{\nu\lambda]}=0$, hence $\omega_{\mu\nu}$ is the K\"ahler form in the space transverse to the vector field $v^\mu$.  

 In even dimensions it is not possible to set all $v^I$, $I\ne0$, everywhere to zero.  However, we can choose an $\eps$ such that $v^{d+1}=\cos\theta$, where $\theta$ is the angle away from the north pole, while $v^I=0$, $I>d+1$.  
 
 In any dimension we can decompose the 16 independent fermions $\Psi$ as
 \be
 \Psi=\sum_{M=1}^9 \Psi_M \tilde\Gamma^M\Gamma^0\Psi+\sum_{m=1}^7\Upsilon_m \nu^m\,,
 \ee
 and so using (\ref{psrel}),
 \be
 \Psi_M=\eps\Gamma_M\Psi\,,\qquad \Upsilon_m=\nu_m\Gamma^0\Psi\,.
 \ee
 Note that 
 \be
 \Psi_0\equiv\eps\Gamma_0\Psi=-\sum_{M=1}^9(\eps\Gamma_M\eps)\Psi^M=-\sum_{M=1}^9v_M\Psi^M\,,
 \ee
 and hence, is not independent.
  The supersymmetry transformations in (\ref{susyos}) then become
 \be\label{susyosr}
 \delta_\eps A_M&=&\Psi_M\nn\\
  \delta_\eps \Psi_M&=&-v^NF_{NM}-[\phi_0,A_M]-\frac{\alpha_I\,d}{4r}\phi_I(\eps\tilde\Gamma_{MI}\Lambda\eps)\,\nn\\
\delta_\eps \Upsilon_m&=&H_m\equiv K_m+\frac{d-3}{r}\nu_m\Lambda\eps\phi_0
+\frac12 F_{MN}(\nu_m\Gamma^{MN0}\eps)-\frac{\al_I\,d}{4r}(\nu_m\Gamma^0\tilde\Gamma^I\Lambda\eps)\phi_I\,,\nn\\
\delta_\eps H_m&=&-v^\mu D_\mu\Upsilon_m-[\phi_0,\Upsilon_m]-(\nu_m\Gamma^\mu\nabla_\mu\nu_n)\Upsilon^n+\frac{d-4}{2r}(\nu_m\Lambda\nu_n)\Upsilon^n\,,
 \ee
 where all $M,N,I\ne0$.  The algebra of course still closes onto a combination of Lie derivatives, gauge transformations, $R$-symmetries and internal $SO(7)$ transformations.

  In the following sections we apply these constructions individually to the case of seven and six dimensions.
 \section{The seven  dimensional theory}
 \label{s-7D}
 
In this section we  specialize to the seven-dimensional sphere. In many respects the seven-dimensional case is analogous
  to the five dimensional construction \cite{Kallen:2012cs, Kallen:2012va},  which is a generalization of   Pestun's original calculation
   on $S^4$  \cite{Pestun:2007rz}. Therefore, we will be brief with the more obvious elements of the calculation  while stressing the  seven dimensional peculiarities. 

\subsection{The 7D cohomological complex}

Our aim in this subsection  is to study the cohomological complex (\ref{susyosr}) in seven dimensions. 

  We first define the fermionic two-forms $\Upsilon_{\mu\nu}$,
\be\label{Upsdef}
\Upsilon_{\mu\nu}\equiv \Upsilon_m (\nu^m \Gamma^{\mu\nu0}\eps)\,.
\ee
In seven dimensions the vector field $v^\mu$ satisfies $v^\mu v_\mu=1$, $v^A=0$, and using (\ref{psrel}) it is straightforward to show that
\be
v^\mu\Upsilon_{\mu\nu}=0\,.
\ee
Hence, the indices of $\Upsilon_{\mu\nu}$ lie in  the six-dimensional horizontal space orthogonal to $v^\mu$.  Using the triality identity and (\ref{psrel}) it is straightforward to show that 
\be
{\omega_\mu}^\sigma\,\omega_{\sigma\nu}=-(g_{\mu\nu}-v_\mu v_\nu)\,.
\ee
By extensively using (\ref{triality}), (\ref{psrel}) and the 10-dimensional chirality of $\epsilon$, which leads to
 \be\label{10dch}
 \eps\Gamma^{\mu\nu\lambda\rho0}\eps=-\frac{1}{2}\eps^{\mu\nu\lambda\rho\sigma_1\sigma_2\kappa}\omega_{\sigma_1\sigma_2}v_\kappa\,,
 \ee
we then observe that
\be
(\iota_v* (\Upsilon\wedge\omega))_{\mu\nu}&=&\Upsilon_{\mu\nu}-\Upsilon^m(\nu_m\Lambda\eps)\omega_{\mu\nu}\nn\\
&=&\Upsilon_{\mu\nu}+\tilde\Upsilon\,\omega_{\mu\nu}\,,
\ee
where
\be
\tilde\Upsilon\equiv\frac{1}{6}\Upsilon_{\lambda\rho}\omega^{\lambda\rho}\,.
\ee
Hence, the $\Upsilon_{\mu\nu}$ can be divided into
\be
\Upsilon_{\mu\nu}=\hat\Upsilon_{\mu\nu}+\tilde\Upsilon \omega_{\mu\nu}\,,
\ee
where $\omega^{\mu\sigma}\hat\Upsilon_{\sigma\nu}=0$.  The $\hat\Upsilon_{\mu\nu}$ satisfy the self-duality condition on the 6D plane {transverse to $v^\mu$} 
\be\label{sd+1}
(\iota_v* (\hat\Upsilon\wedge\omega))_{\mu\nu}=+\hat\Upsilon_{\mu\nu}\,,
\ee
while the piece proportional to $\omega_{\mu\nu}$ satisfies
\be\label{sd+2}
(\iota_v*(\tilde\Upsilon\omega\wedge\omega))_{\mu\nu}=+2\tilde\Upsilon\omega_{\mu\nu}\,.
\ee
There are 6 independent two-forms that are self-dual in six dimensions, and one whose dual is twice the original, namely the K\"ahler form.  Hence, this accounts for the seven independent degrees of freedom in $\Upsilon_m$.  The other 8 two-forms in six dimensions are anti-self-dual.    

The vector field $v^\mu$ generates a free $U(1)$ action  on $S^7$ which corresponds to the Hopf fibration over $\mathbb{C}P^3$, where $\kappa_\mu = g_{\mu\nu} v^\nu$ is the natural connection one-form 
 on this principal bundle. 
  Hence, the $\Upsilon_{\mu\nu}$ are the horizontal two-form on $S^7$.  On the horizontal plane we  can introduce the complex structure  that naturally splits the horizontal forms into 
   $(p,q)$-forms with respect to this complex structure.  The horizontal self-dual two-form $\Upsilon_{\mu\nu}$ can be decomposed to $(2,0)$ and $(0,2)$ forms  and  the $(1,1)$-form which is proportional 
    to $\omega_{\mu\nu}$. Indeed the pull-back of the standard K\"ahler form  on $\mathbb{C}P^3$ to $S^7$  coincides with $\omega_{\mu\nu}$.

We can next show that the $R$-symmetry terms in $\delta_\eps^2\Upsilon_m$ have the right form to make $\delta^2_\eps\Upsilon_{\mu\nu}$ into a Lie derivative plus a gauge transformation.  We first note that
\be\label{trans1}
-\LL^A_v(\Upsilon_m(\nu^m\Gamma^{\mu\nu0}\epsilon))&=&-(\LL^A_v\Upsilon_m)(\nu^m\Gamma^{\mu\nu0}\epsilon)-v^\sigma\nabla_\sigma(\nu^m\Gamma^{\mu\nu0}\epsilon)+v_\sigma\nabla^\mu(\nu^m\Gamma^{\sigma\nu0})+v_\sigma\nabla^\nu(\nu^m\Gamma^{\mu\sigma0})\nn\\
&=&-(\LL^A_v\Upsilon_m)(\nu^m\Gamma^{\mu\nu0}\epsilon)-v^\sigma(\nabla_\sigma\nu^m)\Gamma^{\mu\nu0}\epsilon-\frac{1}{2r}\nu^m\tilde\Gamma^{\mu\nu}\Lambda\eps\nn\\
&&\qquad-\frac{1}{r}(\eps\Gamma_M\tilde\Gamma^\mu\Lambda\eps)\,(\nu^m\tilde\Gamma^{\nu0}\Gamma_M\eps)+\frac{1}{r}(\eps\Gamma_M\tilde\Gamma^\nu\Lambda\eps)\,(\mu^m\tilde\Gamma^{\nu0}\Gamma_M\eps) \,,
\ee
 where $\LL_v^A$ is the covariant version of the Lie derivative, $\LL_v^A = d_A \iota_v + \iota_v d_A$. 
We then take the explicit transformation in (\ref{susyosr}) to write
\be\label{trans2}
(\nu^m\Gamma^{\mu\nu0}\epsilon)\delta_\eps^2\Upsilon_m&=&(\nu^m\Gamma^{\mu\nu0}\epsilon)\left(-\LL^A_v\Upsilon_m-[\phi_0,\Upsilon_m]+\frac{3}{2r}(\nu_m\Lambda\nu_n)\Upsilon^n-(\nu_m\Gamma^\sigma\nabla_\sigma\nu_n)\Upsilon^n\right)\nn
\\
&=&-(\LL^A_v\Upsilon_m+[\phi_0,\Upsilon_m])(\nu^m\Gamma^{\mu\nu0}\epsilon)-\frac{3}{2r}(\nu_n\tilde\Gamma^{\mu\nu}\Lambda\eps)\Upsilon^n\nn\\
&&\qquad-\frac{1}{r}(\eps\Gamma_M\tilde\Gamma^\mu\Lambda\eps)\,(\nu_m\tilde\Gamma^{\nu0}\Gamma_M\eps)\Upsilon^m+\frac{1}{r}(\eps\Gamma_M\tilde\Gamma^\nu\Lambda\eps)\,(\nu_m\tilde\Gamma^{\mu0}\Gamma_M\eps)\Upsilon^m\nn\\
&&\qquad-\frac{1}{2r}(\eps\Gamma_M\tilde\Lambda\Gamma_\sigma\eps)\,(\nu_m\Gamma_M\tilde\Gamma^{\mu\nu0}\Gamma^\sigma\eps)\Upsilon^m\,.
\ee
We can rewrite the last term in (\ref{trans2}) as
\be
-\frac{1}{2r}(\eps\Gamma_M\tilde\Lambda\Gamma_\sigma\eps)\,(\nu_m\Gamma_M\tilde\Gamma^{\mu\nu0}\Gamma^\sigma\eps)\Upsilon^m=-\frac{1}{2r}(\nu_m\tilde\Gamma^{\mu\nu}\Lambda\eps)\Upsilon^m+\frac{3}{2r}(\nu_m\tilde\Gamma^{\mu\nu}\Lambda\eps)\Upsilon^m\,,
\ee
hence, by comparing terms in (\ref{trans1}) and (\ref{trans2}), we see that
\be\label{Upstrans}
-\LL^A_v(\Upsilon_{\mu\nu})-[\phi_0,\Upsilon_{\mu\nu}]=(\nu^m\Gamma^{\mu\nu0}\epsilon)\delta_\eps^2\Upsilon_m\,.
\ee
Therefore, the transformation of the two-form $\Upsilon_{\mu\nu}$ is the negative of the Lie derivative plus the usual gauge transformation.

We can next define the bosonic  three-form fields
\be\label{Phidef}
\Phi_{\mu\nu\lambda}=\sfrac{1}{2}\phi_A(\eps\Gamma_{\mu\nu\lambda}\Gamma^{A0}\eps)\,,
\ee
where we again find that the field is horizontal, satisfying $v^\mu\Phi_{\mu\nu\lambda}=0$.  The Lie derivative of $\Phi_{\mu\nu\lambda}$ is then
\be\label{3-form7}
2\,\LL^A_v \Phi_{\mu\nu\lambda}&=&(\LL_v\phi_A)(\eps\Gamma_{\mu\nu\lambda}\Gamma^{A0}\eps)+\phi_A\LL^A_v(\eps\Gamma_{\mu\nu\lambda}\Gamma^{A0}\eps)\nn\\
&=&(\LL^A_v\phi_A)(\eps\Gamma_{\mu\nu\lambda}\Gamma^{A0}\eps)+\phi_I\left(v^\sigma\nabla_\sigma(\eps\Gamma_{\mu\nu\lambda}\Gamma^{A0}\eps)-3\,v^\sigma \nabla_{[\mu}(\eps\Gamma_{\nu\lambda]\sigma}\Gamma^{A0}\eps)\right)\nn\\
&=&(\LL^A_v\phi_A)(\eps\Gamma_{\mu\nu\lambda}\Gamma^{A0}\eps)+\frac{4}{r}\phi_A\varepsilon_{AB}(\eps\Gamma_{\mu\nu\lambda}\Gamma^{B0}\eps)\nn\\
&=&\left(v^\sigma D_\sigma\phi_A-\frac{4}{r}\varepsilon_{AB}\phi^B\right)(\eps\Gamma_{\mu\nu\lambda}\Gamma^{A0}\eps)\,.
\ee
In (\ref{phit}), the transformation on $\phi_A$ for d=7 becomes
 \be\label{phit7}
 \delta^2_\eps\phi_A&=&-v^\sigma D_\sigma\phi_A-[\phi_0,\phi_A]-\frac{4}{r}\, (\eps\tilde\Gamma_{AB}\Lambda\eps)\,\phi^B\nn\\
&=&-v^\sigma D_\sigma\phi_A-[\phi_0,\phi_A]+\frac{4}{r}\, \varepsilon_{AB}\,\phi^B\,.
 \ee
Comparing (\ref{phit7}) to (\ref{3-form7}), we see that the $R$-symmetry transformation is precisely what is needed to turn the field $\phi_I$ into a natural 3-form under the Lie derivative.

It is also not hard to see that the $\Phi_{\mu\nu\lambda}$ split naturally into (3,0) and (0,3) forms under the complex structure of the horizontal space.  The square of the horizontal dual is $(\iota_v*)^2=-1$, hence a general three-form can split into 10 independent forms  $\Phi^+$ and 10 $\Phi^{-}$ with $\iota_v*\Phi^\pm=\pm i\,\Phi^\pm$.  In this particular case, we can explicitly write the forms  as
\be
\Phi^\pm_{\mu\nu\lambda}=\sfrac{1}{4}(\phi_8\mp i\,\phi_9)\left[(\eps\Gamma_{\mu\nu\lambda}\Gamma^{80}\eps)\pm i\,(\eps\Gamma_{\mu\nu\lambda}\Gamma^{80}\eps)\right]\,.
\ee
The self-dual forms (eigenvalue $+i$)  are either $(3,0)$ or $(1,2)$ forms, while the anti-self-dual forms (eigenvalue $-i$) are $(0,3)$ or $(2,1)$ forms.   By making liberal use of the identities in the appendix, we  can further establish
\be
{\omega_\mu}^{\sigma}\Phi^{\pm}_{\sigma\nu\lambda}=\pm i\,\Phi^{\pm}_{\mu\nu\lambda}\, ,
\ee
which  restricts the forms to be $(3,0)$ or $(0,3)$.  Analogously we can define the superpartner of $\Phi$
\be\label{etadef}
\eta_{\mu\nu\lambda}=\sfrac{1}{2}\Psi_A(\eps\Gamma_{\mu\nu\lambda}\Gamma^{A0}\eps)\,,
\ee
 with  similar properties as $\Phi$. 

Let us now discuss the fixed point locus in the case of 7 dimensions.  
 We take from the first line of  (\ref{FwF}) (prior to the integration by parts),
 \be\label{FwF7}
&& -\frac{1}{4}F_{MN}F_{M'N'}(\eps\Gamma^{MNM'N'0}\eps)=\nn\\
&&\qquad-\frac{1}{4}F_{\mu\nu}F_{\lambda\rho}(\eps\Gamma^{\mu\nu\lambda\rho0}\eps)- F_{\mu\nu}(D_\lambda\phi_A)(\eps\Gamma^{\mu\nu\lambda A0}\eps)+2\omega^{\mu\nu}D_\mu\phi_8D_\nu\phi_{9}-[\phi_8,\phi_9]f\,,\nn\\
 \ee
 where $f\equiv F_{\mu\nu}\omega^{\mu\nu}$.
Using (\ref{10dch}),
 we can rewrite the first term in (\ref{FwF7}) as
 \be\label{FdF}
 -\frac{1}{4}F_{\mu\nu}F_{\lambda\rho}(\eps\Gamma^{\mu\nu\lambda\rho0}\eps)=\frac{1}{2}F_{\mu\nu}(\iota_v*F\wedge \omega)^{\mu\nu}\,.
 \ee
If we decompose $F_{\mu\nu}$ as
\be
F_{\mu\nu}=\hat F_{\mu\nu}+\sfrac{1}{6}f\omega_{\mu\nu}\,,
\ee
then (\ref{FdF}) combines with the gauge kinetic term to give
\be
&&\sfrac{1}{2}F_{\mu\nu}F^{\mu\nu} -\frac{1}{4}F_{\mu\nu}F_{\lambda\rho}(\eps\Gamma^{\mu\nu\lambda\rho0}\eps)\nn\\
&&\qquad\qquad=
\sfrac{1}{4}(\hat F+\iota_v*\hat F\wedge \omega)^2+\sfrac{1}{4}f^2+\sfrac{1}{2}F_{\mu\lambda}{F^\mu}_\rho v^\lambda v^\rho\,.
\ee

We also can see that  only the third term in (\ref{DphiF}) contributes in 7 dimensions, and is found to be
\be
\frac{8}{r}\phi_A v^\lambda D_\lambda \phi_B \varepsilon^{AB}\,.
\ee
Combining this with the scalar terms contributing to (\ref{FwF7}) and $\sfrac12 F_{MN}F^{MN}$, we find that the complete bosonic part of (\ref{bosdPdP}) is
\be\label{cfp}
&&\sfrac{1}{4}(\hat F+\iota_v*\hat F\wedge \omega)^2+\sfrac{1}{4}f^2+\sfrac{1}{2}F_{\mu\lambda}{F^\mu}_\rho v^\lambda v^\rho
-2\,\hat F_{\mu\nu}D_\lambda \Phi^{\mu\nu\lambda}-D_\mu\phi_0D^\mu\phi_0\nn\\
&&\qquad\qquad+D_\mu\phi_A D_\nu\phi_B(g^{\mu\nu}\delta^{AB}+\omega^{\mu\nu}\varepsilon^{AB})-\sfrac12\ve^{AB}[\phi_A,\phi_B]\,f+\frac{8}{r}\phi_A v^\lambda D_\lambda \phi_B \varepsilon^{AB}\nn\\
&&\qquad\qquad
-(K_m+\frac{4}{r}\phi_0(\nu_m\Lambda\epsilon))^2+\frac{16}{r^2}\phi_A\phi^A+\sfrac{1}{2}[\phi_A,\phi_B]^2\nn\\
&&=(\hat F^+_{\mu\nu}-D_\lambda{\Phi_{\mu\nu}}^\lambda)^2+\sfrac{1}{4}(f-\sfrac{1}{6}[\Phi_{\mu\nu\lambda},{\Phi^{\mu\nu}}_\sigma]\omega^{\lambda\sigma})^2+\sfrac{1}{2}F_{\mu\lambda}{F^\mu}_\rho v^\lambda v^\rho
\nn\\
&&\qquad\qquad-D_\mu\phi_0D^\mu\phi_0-(K_m+\frac{4}{r}\phi_0(\nu_m\Lambda\epsilon))^2+\sfrac{1}{6}v^\sigma v^\rho H_{\sigma\mu\nu\lambda}\,{H_\rho}^{\mu\nu\lambda}\,,
\ee
where
\be\label{fpfull7-extra}
\hat F^+&=&\iota_v*(\hat F^+\wedge \omega)\nn\\
H_{\sigma\mu\nu\lambda}&\equiv& D_\sigma \Phi_{\mu\nu\lambda}-D_\mu \Phi_{\sigma\nu\lambda}-D_\nu \Phi_{\mu\sigma\lambda}-D_\lambda \Phi_{\mu\nu\sigma}\,.
\ee
In deriving the second line in (\ref{cfp}) we made use of (\ref{iden2}) and (\ref{iden3}). 
 Next, we Wick rotate $\phi_0 \rightarrow i\phi_0$ and $K_m \rightarrow i K_m$ to make the entire supersymmetric action  positive definite. 
 Hence, the complete fixed point locus is
 \be\label{fpfull7}
 K^m&=&-\frac{4}r\phi_0\,(\nu_m\Lambda\eps)\,,\qquad D_\mu\phi_0=0\nn\\
\hat F^+_{\mu\nu}&=&D_\sigma{\Phi_{\mu\nu}}^\sigma\nn\\
f&=&\sfrac{1}{6}[\Phi_{\mu\nu\lambda},{\Phi^{\mu\nu}}_\sigma]\omega^{\lambda\sigma}\nn\\
v^{\mu}F_{\mu\nu}&=&0\nn\\
v^{\sigma}\,H_{\sigma\mu\nu\lambda}&=&0\,.
\ee
The  $\hat F^+$ in the second line is a (2,0) plus a (0,2) form.  These equations are the seven-dimensional lift of the Hermitian Higgs-Yang-Mills system which previously appeared in \cite{Blau:1997pp} (see \cite{MR2491283} for a nice overview).

 We next  evaluate the supersymmetric action on the localization locus.
The bosonic part of the  Lagrangian (after the Wick rotation $\phi_0 \rightarrow i\phi_0$ and $K_m \rightarrow i K_m$) can be written as
\be\label{LL7}
\LL_{ss}+\LL_{aux}&=&\frac{1}{g_{YM}^2}\Tr\Big[\sfrac{1}{2}(F^+_{\mu\nu})^2+\sfrac{1}{2}(F^-_{\mu\nu})^2+\sfrac{1}{12}f^2+\sfrac{1}{4}(\ve^{AB}[\phi_A,\phi_B])^2+(D_\sigma{\Phi_{\mu\nu}}^\sigma)^2\nn\\
&&\qquad\quad-\omega^{\mu\nu}\ve^{AB}D_\mu\phi_A D_{\nu}\phi_B+v^\mu D_\mu \phi_A v^\nu D_\nu\phi^A+\frac{8}{r^2}\phi_A\phi^A\nn\\
&&\qquad\quad + D_\mu\phi_0D^\mu\phi_0 + \frac{8}{r^2}\phi_0\phi_0 + K^mK_m\Big]
\ee
The topological term in (\ref{FdF}) can be written as
\be
\sfrac{1}{2}F_{\mu\nu}(\iota_v*F\wedge \omega)^{\mu\nu}=\sfrac{1}{2}(F^+_{\mu\nu})^2-\sfrac{1}{2}(F^-_{\mu\nu})^2+\sfrac{1}{6} f^2\,,
\ee
hence, using this and the fixed point locus in (\ref{fpfull7}) we can rewrite (\ref{LL7}) as
\be\label{LL72}
\LL_{ss}+\LL_{aux}&=&\frac{1}{g_{YM}^2}\Tr\Big[2F^+_{\mu\nu}D_\sigma{\Phi_{\mu\nu}}^\sigma-\sfrac{1}{2}F_{\mu\nu}(\iota_v*F\wedge \omega)^{\mu\nu}\nn\\
&&\qquad\quad+\Big(\sfrac{1}{2}\omega^{\mu\nu}F_{\mu\nu}[\phi_A,\phi_B]-\omega^{\mu\nu}D_\mu\phi_A D_{\nu}\phi_B-\frac{6}{r} \phi_Av^\mu D_\mu\phi_B\Big)\ve^{AB}\nn\\
&&\qquad\qquad\qquad +\frac{24}{r^2}\phi_0\phi_0\Big]\,.
\ee
The first term in (\ref{LL72}) is zero after integrating by parts and using the Bianchi identity.  The second line is also zero after  integrating by parts the second term in that line and using (\ref{divom}).  Hence, at the fixed point the Lagrangian reduces to
\be\label{LLfp7} 
\LL_{fp}&=&\frac{1}{g_{YM}^2}\Tr\Big[ \frac{24}{r^2}\phi_0\phi_0-\sfrac{1}{2}F_{\mu\nu}(\iota_v*F\wedge \omega)^{\mu\nu}\Big]\,,
\ee
{\it i.e.} the term written in (\ref{LLfp}) plus an instanton contribution. The integral over (\ref{LLfp7}) can be rewritten 
 as follows
\be\label{integral-LLfp7}
 \int \LL_{fp} \sqrt{g} ~d^7 x &=& \frac{1}{g_{YM}^2}\Big[ \frac{24}{r^2}\int \Tr (\phi_0^2) {\rm vol}_g + r \int \Tr (F\wedge F \wedge \kappa \wedge d\kappa) \Big ]~,
\ee
 where we have used (\ref{kappa-omega}) and the relation $\iota_v (*\omega_p) = (-1)^p (\kappa \wedge \omega_p)$ for $p$-form
  $\omega_p$.

 \subsection{Calculation of determinants}

 Let us summarize in a coordinate free way the supersymmetry  transformations from the last section and calculate the one loop determinant around 
  the trivial connection $A=0$.   The present  discussion is very similar to the 5D case \cite{Kallen:2012cs, Kallen:2012va} and we refer the reader there for some basic definitions from
  contact geometry.  
 
 Using the projectors  $\kappa \wedge i_v$ and $(1- \kappa\wedge i_v)$ we can decompose the differential form
  into vertical and horizontal parts 
  $$ \Omega^\bullet (S^7) = \Omega_V^\bullet (S^7) \oplus \Omega_H^\bullet (S^7)~.$$
 Furthermore using the complex structure on $\ker k$  the horizontal forms can be decomposed further into $(p,q)$-forms, $\Omega_H^{(p,q)}(S^7)$. 
 
 Using the field redefinitions from the last subsection and after the Wick rotation of $\phi_0$
   the supersymmetry transformations (\ref{susyosr}) can be rewritten in terms of even and odd
  differential forms as follows
 \be\label{7D-summary-tr}
&& \delta_\epsilon A = \psi~,\nn\\
 && \delta_\epsilon \psi = - i_v F - i d_A \phi_0,\nn \\
 && \delta_\epsilon \phi_0 = i \, i_v \psi~,\nn \\
 && \delta_\epsilon \Phi = \eta~,\\
 && \delta_\epsilon \eta = - {\cal L}_v^A \Phi - i [\phi_0, \Phi]~,\nn \\
 && \delta_\epsilon \Upsilon = H ~,\nn \\
 && \delta_\epsilon H = - {\cal L}^A_v \Upsilon - i [\phi_0, \Upsilon]~, \nn
 \ee
  where $d_A$ is the de Rham differential coupled to the connection $A$  and ${\cal L}_v^A = i_v d_A + d_A i_v$ is a gauge covariant 
   Lie derivative.  Here $A$ is a connection and all other fields are in the adjoint representation.
   $\Psi_M$ goes to an odd 1-form $\psi$ and to an odd horizontal $(3,0) + (0,3)$-form $\eta$, $\phi_0$ is an even scalar, and $\Phi$ is an even  horizontal
    $(3,0) + (0,3)$-form.
    $\Upsilon$ is an odd horizontal self-dual form (i.e., a horizontal $(2,0)+(0,2)$ plus a $(1,1)$ part proportional to $\omega$) and $H$ is an
    even horizontal self-dual form. 
   
   The original supersymmetry dictates that the vector $v$ is along the Hopf fibers if we think of $S^7$ as $S^1$-fibration 
    over $\mathbb{C}P^3$. However once the transformations are written in the form (\ref{7D-summary-tr}) we can choose $v$ to be 
     any combinations of four $U(1)$s acting on $S^7$.  The seven-sphere  is defined by the following equation in $\mathbb{C}^4$
       \be
      |z_1|^2 + |z_2|^2 + |z_3|^2 + |z_4|^2 =1~.  
     \ee
  There is the following $T^4$ action on $\mathbb{C}^4$
  \be
   z_i~\rightarrow~ e^{i\alpha_i} z_i~,~~~~~i=1,2,3,4,
  \ee
   which descends to $S^7$. Let us denote by $e_i$ the vector field on $S^7$ for the corresponding $U(1)$ action. Then 
    in general we can consider $v = \sum\limits_{i=1}^4 \omega_i e_i$ corresponding to a so-called squashed sphere. 
     If $\omega_i=1$, $i=1,2,3,4$ then this $v$ corresponds to the $S^1$ Hopf fibration of a round sphere. Allowing $v$ to be general 
      toric will help us  conjecture the full answer for the partition function. 
 
  Before we start to calculate the path integral, we have to gauge fix the theory. This is a standard topic in localization and thus we will be brief.
     We follow closely the original work \cite{Pestun:2007rz} by Pestun. We  introduce the standard ghosts $c, \bar{c}$ and Lagrangian multipler 
      $b$. Moreover we have to introduce the even zero modes $(a_0, \bar{a}_0, b_0)$ and the odd zero modes $(c_0, \bar{c}_0)$.
        We also introduce the standard BRST transformation and  combine it with the supersymmetry $\delta_\epsilon$ into a new 
         transformation $Q$. This can be done in the standard way  \cite{Pestun:2007rz}. For the actual calculation, it is important that 
          $Q^2 =- {\cal L}_v - i G_{a_0}$ and $Q$ acts as an equivariant differential on the supermanifold with even coordinates $(A, \Phi, \bar{a}_0, b_0)$
           and odd coordinates $(\Upsilon, c, \bar{c})$. Here $G_{a_0}$ is the gauge transformation with  constant parameter $a_0$. 
            We  impose the Lorentz gauge and the parameter $a_0$ is identified with $\phi_0$. 
 
 The localization locus is given by equations (\ref{fpfull7}). If we choose the isolated solution  $A=0, \Phi=0$ and $\phi_0 = {\rm constant}= r^{-1}\sigma$
  then the 1-loop contribution around this locus corresponds to the full perturbative answer. Thus the full perturbative answer is given by
  \be
   \int\limits_{\mathbf{g}} d\sigma  ~e^{ - \frac{8\pi^4 r^3}{g_{YM}^2} {\rm Tr}(\sigma^2)} \frac{ \sqrt{ det_{\Omega_H^{(2,0)}} (Q^2)  det_{\Omega_H^{(0,2)}} (Q^2) (det_{\Omega_H^{(0,0)}} (Q^2))^3 }}{ \sqrt{ det_{\Omega^1} (Q^2) det_{\Omega_H^{(3,0)}}(Q^2) det_{\Omega_H^{(0,3)}} (Q^2)
    (det_{H^0} (Q^2))^2}}~,
  \ee
  where the denominator comes from integration over the even coordinates  $(A, \Phi, \bar{a}_0, b_0)$ and the numerator from the integration over the odd coordinates $(\Upsilon, c, \bar{c})$.
   Here all forms are Lie algebra valued. 
   Decomposing $\Omega^1 = \Omega^0 k \oplus \Omega_H^{(1,0)} \oplus \Omega_H^{(0,1)}$, cancelling some pieces and taking the square root (we ignore the phase
    of the determinants, we are interested in the absolute value only) we are left to calculate the following  superdeterminant 
    \be\label{super-det7D}
   {\rm sdet}_{\Omega^{(\bullet,0)}_H}(-{\cal L}_v - i [\sigma, ~])~.
   \ee

   Following \cite{Schmude:2014lfa} and \cite{Qiu:2014oqa}   the calculation of the superdeterminant (\ref{super-det7D}) for generic toric $v$ can be reduced to counting
 holomorphic functions on the metric cone over $S^7$,  which is just $\mathbb{C}^4$.  Let us sketch the logic here and later  we present a different approach for 
     the calculation of (\ref{super-det7D}) for the case of  the round sphere. Since the complex structure on $\ker k$ is integrable, we can introduce the horizontal 
      Dolbeault differential $\partial_H$. Since ${\cal L}_v$ commutes with   $\partial_H$ then the calculation of the superdeterminant over $\Omega^{(\bullet,0)}_H$  descends 
       to the calculation of the same superderminant over the cohomology $H_{KR}^{(\bullet,0)}$ with respect to $\partial_H$, the so-called  Kohn-Rossi cohomology. 
        The calculation of the Kohn-Rossi cohomology and the decomposition into eigenspaces for $T^4$  can be reduced to the study of  homolomorphic functions on 
         the metric cone, $\mathbb{C}^4$. The seven dimensional calculation is a straightforward generalization of the 5D case and further details can be found in 
      \cite{Schmude:2014lfa} and \cite{Qiu:2014oqa}. The final answer is given by  the following matrix model
  \be\label{final-S4MM}
 \int\limits_{\mathbf g} d\sigma  && e^{- \frac{8\pi^4 r^3\rho}{g_{YM}^2} {\rm Tr}(\sigma^2)} {\rm det}_{adj}~  {\rm sdet}_{\Omega^{(\bullet,0)}_H}(-  {\cal L}_v - i [\sigma, ~]) = \nn \\
&& \int\limits_{\mathbf{t}} d\sigma~ e^{- \frac{8\pi^4 r^3\rho}{g_{YM}^2} {\rm Tr}(\sigma^2)} \prod\limits_{\alpha} S_4 (i \langle \sigma, \alpha \rangle;  \omega_1, \omega_2, \omega_3, \omega_4)~,
   \ee
   where $\alpha$ are the roots for the Lie algebra $\mathbf{g}$, $\mathbf{t}$ is the Cartan subalgebra,   $v = \sum\limits_{i=1}^{4}\omega_i e_i$ where the $e_i$ correspond to the $T^4$-action on $S^7$, and
    $\rho$ is the ratio of the volume of the squashed sphere to the round sphere. 
   The quadruple sine  is defined as follows (see \cite{koyama2005} for further details)
\bea
 S_4 (x; \omega_1, \omega_2, \omega_3, \omega_4) = \frac{\prod\limits_{i,j,k,l=0}^\infty (i\omega_1 + j \omega_2 + k \omega_3 + l \omega_4 +x)}{\prod\limits_{i,j,k,l=1}^\infty (i\omega_1 + j \omega_2 + k \omega_3 + l \omega_4 - x)}\label{def-S4-general}
\eea
 and the triple sine as 
 \be
  S_3 (x; \omega_1, \omega_2, \omega_3) =  \prod\limits_{i,j,k=0}^\infty (i\omega_1 + j \omega_2 + k \omega_3 + x) \prod\limits_{i,j,k,l=1}^\infty (i\omega_1 + j \omega_2 + k \omega_3  - x)~.
\ee
 For the convergence of $S_4$ one has to 
    require that ${\rm Re} (\omega_i) \geq 0$ (the same is required for  $v$ to be a Reeb vector for some toric contact structure). 
     The contribution of the numerator in $S_4$ comes from counting  in $ H^{(0,0)}_{KR} (S^7) \equiv H^0 ({\cal O} (\mathbb{C}^4))$  
      and the denominator  from counting in {$H^{(3,0)}_{KR} (S^7) \equiv (H^{(0,0)}_{KR} (S^7))^*$} (with some shifts in the allocation of $U(1)$-charges). 
      All other cohomologies
       vanish. In this section we assume that all infinite products are appropriately regularized. The details of the regularization will be discussed in section \ref{s-matrix}. 
       
    Thus the full equivariant answer for the  perturbative partition function for maximally supersymmetric Yang-Mills on squashed $S^7$ is given by  the matrix model (\ref{final-S4MM}).
     The full answer will contain the expansions  around every non-trivial solution of the hermitian Higgs-Yang-Mills system (\ref{fpfull7-extra}) and (\ref{fpfull7}).
      In the next subsection we conjecture the form of the full partition function. 
      
Now  let us concentrate on the case of the round sphere where $\omega_i =1$ for $i=1,2,3,4$. This case is  important for our study of the matrix model 
 and moreover yields an alternative derivation of the determinants. 
    We use the  short hand notation, $S_4(x) = S_4(x; 1,1,1,1)$ and $S_3(x)=S_3(x; 1,1,1)$.  We then derive
     some identities for $S_4(x)$. We have the following 
      \bea
       \prod\limits_{i,j,k,l=0}^\infty  (i + j + k + l + x) = \prod\limits_{t=0}^\infty \prod\limits_{j=0}^{t} \prod\limits_{k=0}^{t-j}
       \prod\limits_{l=0}^{t-j -k} (t+x)  ~, 
      \eea
    where $t= i+ j + k + l \geq 0 $ and so $i = t - j-k -l \geq 0$, $t-j-k \geq l \geq 0$ etc.  We then have
   \bea
     \sum\limits_{j=0}^{t} \sum\limits_{k=0}^{t-j}
       \sum\limits_{l=0}^{t-j -k} 1 = \frac{1}{6}t^3 + t^2 + \frac{11}{6} t + 1= \frac{(t+1)(t+2)(t+3)}{6}~,
   \eea 
    where we  used the  identities
    \bea
     \sum\limits_{k=1}^n k = \frac{n(n+1)}{2}= \frac{n^2 +n}{2}~,~~~~~\sum\limits_{k=1}^n k^2 = \frac{n(n+1)(2n+1)}{6}=
     \frac{2n^3 + 3n^2 + n}{6}~. 
    \eea      
    Thus we get 
    \bea
     \prod\limits_{i,j,k,l=0}^\infty  (i + j + k + l + x) =  \prod\limits_{t=0}^\infty (t+x)^{\frac{1}{6}t^3 + t^2 + \frac{11}{6} t + 1}~.
    \eea
 Analogously, we find
    \bea
      \prod\limits_{i,j,k,l=1}^\infty  (i + j + k + l - x) =   \prod\limits_{t=0}^\infty (t + 4 - x)^{\frac{1}{6}t^3 + t^2 + \frac{11}{6} t + 1}~.
    \eea
 This leaves the following expression for $S_4(x)$
    \bea
     S_4(x) = \frac{\prod\limits_{t=0}^\infty (t+x)^{\frac{1}{6}t^3 + t^2 + \frac{11}{6} t + 1}}{\prod\limits_{t=0}^\infty (t + 4 - x)^{\frac{1}{6}t^3 + t^2 + \frac{11}{6} t + 1}} = x \prod\limits_{t=1}^\infty \frac{(t+x)^{\frac{1}{6}t^3 + t^2 + \frac{11}{6} t + 1}}{(t-x)^{\frac{1}{6}t^3 - t^2 + \frac{11}{6} t - 1}}\label{S4-round}\,.
    \eea
    The appearance of the expression (\ref{S4-round}) can be explained through the decomposition of the horizontal differential forms
     in terms of the eigenfunctions of the operator ${\cal L}_v$, where $v$ corresponds to the Hopf fibration.  
  For the round sphere where $v$ goes along an $S^1$-fiber we can  Fourier expand 
    the horizontal forms
  \be
   \Omega_H^{(\bullet, 0)} (S^7, \mathbf{g}) =  \Omega^{(\bullet, 0)} (\mathbb{C}P^3, \mathbf{g}) \bigoplus \Big ( \bigoplus\limits_{t\neq 0}   \Omega^{(\bullet, 0)} (\mathbb{C}P^3, O(t) \otimes \mathbf{g}) \Big ) ~, 
  \ee   
  where $O(1)$ is line bundle associated to the $S^1$-fibration. Up to  some numerical factor corresponding to the normalization,  $t$ is  the eigenvalue of ${\cal L}_v$.   
   To control the cancelation between the numerator and denominator of (\ref{super-det7D}) we have to use the index theorems for the Dolbeault complex twisted by an $O(t)$ bundle. 
    All necessary information is encoded in  this twisted cohomology, 
        \bea
   && H^0 (\mathbb{C}P^3, O(t)) = \frac{(t+1)(t+2)(t+3)}{6}~,~~~t \geq 0~,\\
   && H^0 (\mathbb{C}P^3, O(t)) =0~, ~~~t <0\\
   && H^{(0,3)} (\mathbb{C}P^3, O(t)) = H^{(0,0)} (\mathbb{C}P^3, O(-4) \otimes O(-t))^*
    \eea 
    and thus we arrive at (\ref{S4-round}). In (\ref{S4-round}) the linear term in front of the whole expression comes from the Vandermonde determinant
     (changing the integration from the whole algebra to the Cartan) and the rest comes from the determinant calculation.  The calculation here is  similar to the one-loop calculation for the round $S^5$.  More details about this case can be found in \cite{Kallen:2012cs}. 
         
   Indeed the matrix model for the round $S^7$ can be simplified further due to the reflection symmetry, $\alpha \rightarrow - \alpha$,  of the product over the roots.  
   Thus using the expression (\ref{S4-round}) we get
  \bea
   \prod_\alpha S_4 (i \langle \sigma, \alpha \rangle ) =   \prod_\alpha (i \langle a , \alpha\rangle) \prod\limits_{t=1}^\infty
    (t+ i \langle \sigma, \beta \rangle)^{2t^2 + 2}~.
  \eea  
  Using the following expression for the triple sine
  \bea
   S_3(x) = x \prod\limits_{t=1}^\infty (t+x)^{\frac{1}{2} t^2 + \frac{3}{2} t +1}  \prod\limits_{t=1}^\infty (t-x)^{\frac{1}{2} t^2 - \frac{3}{2} t +1}\,,
  \eea
and for the ordinary sine 
\bea
\sin (\pi x) = x \prod\limits_{t=1}^\infty (t+x) \prod\limits_{t=1}^\infty (t-x)\,,
\eea
we find 
\bea
    \prod_\alpha S_4 ( i \langle \sigma, \beta\rangle ) =   \prod_\alpha \frac{\Big (S_3 (i \langle \sigma, \alpha \rangle )\Big )^2}{\sin ( i\pi
     \langle \sigma, \alpha \rangle )}~.\label{S4-S3}
\eea
Therefore, 
the matrix model (\ref{final-S4MM}) for maximally supersymmetric Yang-Mills on the round $S^7$ can be rewritten as follows
\be\label{7Dpertpf}
Z_{7D}= \int\limits_{\mathbf{t}} d\sigma~ e^{- \frac{8\pi^4 r^3}{g_{YM}^2} {\rm Tr}(\sigma^2)}  
\prod_\alpha \frac{\Big (S_3 (i \langle \sigma, \alpha \rangle )\Big )^2}{\sin ( i\pi
     \langle \sigma, \alpha \rangle )}~.\label{7D-matrix-model}
\ee
  We will discuss the properties of this matrix model in section \ref{s-matrix}. 
 
 We conclude this subsection with the following remark. The supersymmetric Yang-Mills action is not the only observable which 
  is invariant under the supersymmetry transformation (\ref{7D-summary-tr}). One can construct a 7D supersymmetric Chern-Simons 
  action \cite{Kallen:2012cs}. The only difference in the matrix model  will be the appearance of a   $\Tr(\sigma^4)$-term in the exponent. 
   While the supersymmetrization of the $\Tr (F \wedge F\wedge F \wedge \kappa)$ term will lead to a $\Tr(\sigma^3)$-term in the matrix
    model. Almost nothing is known about the dynamics of higher dimensional Chern-Simons theories and thus it will be interesting 
     to study these models 
     through localization. 
 
 \subsection{Factorization and the full answer}
 
 In the previous subsection we have calculated the full perturbative partition function (\ref{final-S4MM}) for $S^7$, in particular 
  the full equivariant version of the answer. In order to derive the full answer we have to expand over all non-trivial solutions of 
   the system (\ref{fpfull7-extra}) and (\ref{fpfull7}). At the moment we do not know how to do it from  first principles. However 
    one would expect that only $T^4$-invariant solutions of those equations will actually contribute to the final answer. For a generic 
     $v$ (which is a combination of the four $U(1)$s) there will be only four closed orbits and thus the solutions will tend to concentrate around 
      these orbits.  {We may be forced to introduce the term $\Tr (F \wedge F\wedge F \wedge \kappa)$ to measure the contributions 
       of these configurations since due  to a simple scaling argument the 7D Yang-Mills action will be zero on such configurations.  In principle 
        the PDEs  (\ref{fpfull7-extra}) and (\ref{fpfull7}) may have other  extended solutions (e.g., membrane-like solutions), but since there are no 
         non-trivial 3-cycles that they can wrap, we believe that they do not contribute to the path integral.}
            Thus this logic may allow us to conjecture the full answer for $S^7$. In this we follow the logic applied to the localization 
       results for  5D supersymmetric gauge theories \cite{Qiu:2013aga} and \cite{Qiu:2014oqa}. 
 
 Let us start by introducing the special functions and discuss their relation to the quadruple sine.  For the regions ${\rm Im} ~\epsilon_i  >  0$
  we define the following special functions 
 \be
  (z| \epsilon_1, \epsilon_2, \epsilon_3) = \prod_{s,n,k=0}^\infty (1 -e^{2\pi i z} e^{2\pi i \epsilon_1 s} e^{2\pi i \epsilon_2 n}e^{2\pi i \epsilon_3 k})
 \ee
  and for the other regions it can be defined as follows
  \be
  (z| \epsilon_1, \epsilon_2, \epsilon_3) = \begin{cases}
 \prod\limits_{s,n,k=0}^\infty (1 -e^{2\pi i z} e^{-2\pi i \epsilon_1 (s+1)} e^{2\pi i \epsilon_2 n}e^{2\pi i \epsilon_3 k})^{-1}~, ~~{\rm Im}~\epsilon_1 <0,  {\rm Im}~\epsilon_2 > 0, {\rm Im}~\epsilon_3 > 0&\\
  \prod\limits_{s,n,k=0}^\infty (1 -e^{2\pi i z} e^{-2\pi i \epsilon_1 (s+1)} e^{-2\pi i \epsilon_2 (n+1)}e^{2\pi i \epsilon_3 k})~,~~{\rm Im}~\epsilon_1 <0,  {\rm Im}~\epsilon_2 < 0, {\rm Im}~\epsilon_3 > 0 &\nonumber\\
\prod\limits_{s,n,k=0}^\infty (1 -e^{2\pi i z} e^{-2\pi i \epsilon_1 (s+1)} e^{-2\pi i \epsilon_2 (n+1)}e^{-2\pi i \epsilon_3 (k+1)})^{-1} ~,~~{\rm Im}~\epsilon_1 <0,  {\rm Im}~\epsilon_2 < 0, {\rm Im}~\epsilon_3 < 0&
  \end{cases}
 \ee
  The quadruple sine  was defined in (\ref{def-S4-general}) with $\omega_i$  assumed  positive real. However 
    $S_4 (x; \omega_1, \omega_2, \omega_3, \omega_4)$ is also well-defined for complex $\omega_i \in {\mathbb{C}}$ as long as ${\rm Re}~ \omega_i \geq 0$. If we allow non-zero imaginary parts for $\omega_i$ then 
 the  quadruple sine admits the following factorization formula \cite{MR2101221}
 \be\label{factor-S4}
 S_4 (x; \omega_1, \omega_2, \omega_3, \omega_4)&=&e^{\frac{\pi i}{24}B_{44}(x|\vec{\omega})}\left (\frac{x}{\omega_1}| \frac{\omega_2}{\omega_1}, \frac{\omega_3}{\omega_1}, 
 \frac{\omega_4}{\omega_1} \right ) \left (\frac{x}{\omega_2}| \frac{\omega_1}{\omega_2}, \frac{\omega_3}{\omega_2}, 
 \frac{\omega_4}{\omega_2} \right ) \times \nonumber \\
&&\left (\frac{x}{\omega_3}| \frac{\omega_1}{\omega_3}, \frac{\omega_2}{\omega_3}, 
 \frac{\omega_4}{\omega_3} \right )
\left (\frac{x}{\omega_4}| \frac{\omega_1}{\omega_4}, \frac{\omega_2}{\omega_4}, 
 \frac{\omega_3}{\omega_4} \right )~,
 \ee
  where $B_{44}(x|\vec{\omega})$ is a multiple Bernoulli polynomial, a  fourth order polynomial in $x$ whose  concrete form is not 
   important for us.  
 
 The perturbative partition function for $\mathbb{C}^3 \times S^1$ for the maximally supersymmetric theory 
  is given by the following expression  \cite{MR2491283}
  \be\label{pert-flat-7D}
  Z_{pert}^{7D}(\sigma;\beta, \epsilon_1, \epsilon_2, \epsilon_3) =  \prod_\alpha ( \beta \langle \sigma, \alpha \rangle | \beta\epsilon_1, \beta\epsilon_2, \beta \epsilon_3)~,
  \ee
  where we have ignored the $\sigma$-independent part and assumed ${\rm Im}~ \epsilon_i < 0$.  Here $\beta$ is the radius of $S^1$. 
   Combining the factorization of the quadruple sine (\ref{factor-S4}) and the perturbative answer  (\ref{pert-flat-7D}) on $\mathbb{C}^3 \times S^1$
    we see that upon  analytical continuation in the $\omega_i$ the perturbative 1-loop result is factorized into four pieces corresponding to 
     $\mathbb{C}^3 \times S^1$ with the following identifications $\beta = 1/\omega_1, \epsilon_1=\omega_2, \epsilon_2 =\omega_3, 
      \epsilon_3=\omega_4$ plus cyclic permutations on   $\omega_i$. 
 
 The instanton partition function on $\mathbb{C}^3 \times S^1$ will count the 6D version of the equations (\ref{fpfull7-extra}) and (\ref{fpfull7}). 
  For $U(N)$ the instanton corrections can be written in terms of colored three dimensional partitions  \cite{MR2491283}. Let us denote  the instanton partition function on $\mathbb{C}^3 \times S^1$ by 
 \be\label{7Dinstpf}
  Z_{inst}^{7D}(\sigma;\beta, \epsilon_1, \epsilon_2, \epsilon_3, q)\,,
 \ee
 where $q$ is the instanton counting parameter.  The explicit form of (\ref{7Dinstpf})
 is not important for us here.
   On $S^7$ it is natural to conjecture that the full instanton 
  answer will be given by four copies of $Z_{inst}^{7D}$ with the same identification of parameters as in the factorization of 
   the perturbative answer. 
  Thus the full answer can be written as an integral over four copies of the flat answer for $\mathbb{C}^3 \times S^1$ in the following manner
 \be
Z^{full}_{7D}= \int\limits_{\mathbf{t}} d\sigma~ e^{- \frac{8\pi^4 r^3\rho}{g_{YM}^2} {\rm Tr}(\sigma^2)}  
  Z_{pert}^{7D}(i\sigma;\frac{1}{\omega_1}, \omega_2, \omega_3, \omega_4)   Z_{inst}^{7D}(i\sigma;\frac{1}{\omega_1}, \omega_2, \omega_3, \omega_4, q)    ({\rm cyclic ~in} ~\omega_{1,2,3,4})\nonumber\\
\ee
 It would be interesting to derive this formula from  first principles. 
 
 \section{Six dimensional theory}
 \label{s-6D}
 
 In this section we derive the perturbative partition function on $S^6$ for the maximally supersymmetric Yang-Mills theory. The calculation is 
similar to the case of $S^4$. The case of a  round $S^4$ was studied in  \cite{Pestun:2007rz} and later generalized to a 
squashed $S^4$ (i.e., the fully equivariant answer on $S^4$) in \cite{Hama:2012bg}. 
    Since in many aspects our calculation for the determinants is 
    similar to \cite{Pestun:2007rz} and \cite{Hama:2012bg}, we will be brief in  this part of     our 
exposition.

\subsection{Localization locus}

Analogous to  the supersymmetry on  $S^4$, the supersymmetry transformations on $S^6$ do not admit any simple cohomological description in terms 
 of differential forms. Thus we will work with the original fields used in section \ref{s-offshell-susy}.  We start by analyzing the localization locus for the theory on $S^6$. 

To find the localization locus, we can take (\ref{bosdPdP}) and (\ref{FwFphiF|}), specializing to six dimensions.   In this case we find that  
\bea\label{6dloc}
\delta_\eps\Psi\, \overline{\delta_\eps\Psi}&=& \sfrac12 F_{\mu\nu}F^{\mu\nu}-D_p\phi_0D^p\phi_0+D_p\phi_A D^p\phi^A+D_\mu\phi_7D^\mu\phi_7-[\phi_0,\phi_A]^2\nn\\
&&\qquad-\sfrac{1}{4}F_{\mu\nu}F_{\lambda\rho}(\eps\Gamma^{\mu\nu\lambda\rho0}\eps)+2\beta\phi_A D_p\phi_B\varepsilon^{AB}v^p-aD_\mu\phi_I F_{\lambda\rho}(\eps\Gamma^{\mu I\lambda\rho0}\eps)\nn\\
&&\qquad-2\beta(3\!-\!4a)\phi_7 F_{\mu\nu}(\eps\Gamma^{\mu\nu789}\eps)-(K^m\!+\!6\beta\phi_0(\nu_m\Lambda\eps))^2+4\beta^2(9\phi_A\phi^A\!+\!\phi_7\phi_7)\,,\nn\\
\eea
where the index $p$ runs from 1 to 7 and the index $A$ from 8 to 9.  In six dimensions we have that $\eps\Gamma^{\mu\nu\lambda\rho0}\eps=-\sfrac12 \ve^{\mu\nu\lambda\rho\sigma\kappa}\tilde\omega_{\sigma\kappa}$, where $\tilde\omega_{\sigma\kappa}\equiv\eps\Gamma_{\sigma\kappa}\Gamma^{789}\eps$.  We then assume that $\eps_s$ defined in (\ref{GKS}) satisfies $\Gamma^{07}\eps_s=-\eps_s$, in which case we can decompose $\tilde\omega_{\sigma\kappa}$ into
\be\label{omrel}
\tilde\omega_{\sigma\kappa}=\cos^2\sfrac12\theta\,\omega^{\bf N}_{\sigma\kappa}+\sin^2\sfrac12\theta\,\omega^{\bf S}_{\sigma\kappa}\,,
\ee
where $\theta$ is the angle away from the north pole of the six-sphere.  The $\omega^{\bf N,S}_{\sigma\kappa}$ are given by \mbox{$\omega^{\bf N}_{\sigma\kappa}=\eps_s\Gamma_{\sigma\kappa}\Lambda\eps_s$} and $\omega^{\bf S}_{\sigma\kappa}=-(\beta x)^{-2}\tilde\eps_c\Gamma_{\sigma\kappa}\Lambda\tilde\eps_c$. 
 After some algebra one can show that the two-forms are $\omega^{\bf N}_{\mu\nu}= g_{\mu\rho} (J^{\bf N})^\rho_\nu$ and $\omega^{\bf S}_{\mu\nu}= g_{\mu\rho} (J^{\bf S})^\rho_\nu$ 
 with $J^{\bf N,S}$ being two almost complex structures and $g$ is the round 
  metric which is hermitian with respect to 
  both almost complex structures. 
  We also have the closed form
\be
\omega_{\sigma\kappa}=\cos^2\sfrac12\theta\,\omega^{\bf N}_{\sigma\kappa}-\sin^2\sfrac12\theta\,\omega^{\bf S}_{\sigma\kappa}\,.
\ee
With these definitions we can then regroup  (\ref{6dloc}) as 
\be\label{6dloc2}
\delta_\eps\Psi\, \overline{\delta_\eps\Psi}&=&\left(D_\mu\phi_7-\frac{a}{2}F_{\lambda\rho}(\eps{\Gamma_\mu}^{7\lambda\rho0}\eps)\right)^2+a^2\,v^\lambda v^\sigma F_{\mu\lambda}{F^\mu}_\sigma\nn\\
&&+\cos^2\sfrac12\theta\left((3\!-\!4a)^2\left(\sfrac12 F_{\mu\nu}F^{\mu\nu}+ \sfrac12 F_{\mu\nu}*\!(F\wedge\omega^{\bf N})^{\mu\nu}\right)-2\beta(3\!-\!4a)\beta F_{\mu\nu}\omega^{{\bf N}\mu\nu}\phi_7+4\beta^2\phi_7^2\right)\nn\\
&&+\cos^2\sfrac12\theta\left(1-2a^2\sin^2\sfrac12\theta-(3\!-\!4a)^2\right)\left(\sfrac12 F_{\mu\nu}F^{\mu\nu}+ \sfrac12 F_{\mu\nu}*\!(F\wedge\omega^{\bf N})^{\mu\nu}\right)\nn\\
&&+\sin^2\sfrac12\theta\left((3\!-\!4a)^2\left(\sfrac12 F_{\mu\nu}F^{\mu\nu}+ \sfrac12 F_{\mu\nu}*\!(F\wedge\omega^{\bf S})^{\mu\nu}\right)-2\beta(3\!-\!4a)\beta F_{\mu\nu}\omega^{{\bf S}\mu\nu}\phi_7+4\beta^2\phi_7^2\right)\nn\\
&&+\sin^2\sfrac12\theta\left(1-2a^2\cos^2\sfrac12\theta-(3\!-\!4a)^2\right)\left(\sfrac12 F_{\mu\nu}F^{\mu\nu}+ \sfrac12 F_{\mu\nu}*\!(F\wedge\omega^{\bf S})^{\mu\nu}\right)\nn\\
&&+D_{\hat p}\phi_AD^{\hat p}\phi^A+(v^p D_{ p}\phi_A+\beta\phi^B\eps_{AB})^2+35\beta^2\phi_A\phi^A\nn\\
&&+D_p\phi_0D^p\phi_0+(K^m+6\beta\phi_0(\nu_m\Lambda\eps))^2+[\phi_0,\phi_A]^2\,,
\ee
where we have Wick rotated $\phi_0$ and $K^m$, used $\hat p$ to indicate components orthogonal to $v^p$, and used the identity in (\ref{6dsq}).   

We then  observe that the 15 components of $F_{\mu\nu}$ can be decomposed with respect to either $\omega^{\bf N,S}_{\mu\nu}$ in the form
\be
F_{\mu\nu}&=& F_{\mu\nu}^{{\bf N}+}+F_{\mu\nu}^{{\bf N}-}+\sfrac16 f^{\bf N}\omega^{\bf N}_{\mu\nu}\nn\\
&=&  F_{\mu\nu}^{{\bf S}+}+ F_{\mu\nu}^{{\bf S}-}+\sfrac16  f^{\bf S}\omega^{\bf S}_{\mu\nu}
\ee
where the six independent components of $F^{{\bf N,S}+}$ and the eight independent components of $F^{{\bf N,S}-}$ satisfy
\be
F^{{\bf N,S}\pm}=\pm *(F^{{\bf N,S}\pm}\wedge\omega^{\bf N,S})\,.
\ee
We can then write (\ref{6dloc2}) as
\be\label{6dloc3}
&&\delta_\eps\Psi\, \overline{\delta_\eps\Psi}=\nn\\
&&\left(D_\mu\phi_7-\frac{a}{2}F_{\lambda\rho}(\eps{\Gamma_\mu}^{7\lambda\rho0}\eps)\right)^2+a^2\,v^\lambda v^\sigma F_{\mu\lambda}{F^\mu}_\sigma\nn\\
&&+\cos^2\sfrac12\theta\left(\left(\phi_7-\frac{3\!-\!4a}{2}f^{\bf N}\right)^2 +\sfrac14\left(1\!-\!2a^2\sin^2\sfrac12\theta\!-\!(3\!-\!4a)^2\right)(f^{\bf N})^2+\left(1\!-\!2a^2\sin^2\sfrac12\theta\right) F^{\bf N+}_{\mu\nu}F^{{\bf N}+\mu\nu}\right)\nn\\
&&+\sin^2\sfrac12\theta\left(\left(\phi_7-\frac{3\!-\!4a}{2}f^{\bf S}\right)^2 +\sfrac14\left(1\!-\!2a^2\cos^2\sfrac12\theta\!-\!(3\!-\!4a)^2\right)(f^{\bf S})^2+\left(1\!-\!2a^2\cos^2\sfrac12\theta\right) F^{\bf S+}_{\mu\nu}F^{{\bf S}+\mu\nu}\right)\nn\\
&&+D_{\hat p}\phi_AD^{\hat p}\phi^A+(v^p D_{ p}\phi_A+\beta\phi^B\eps_{AB})^2+35\beta^2\phi_A\phi^A\nn\\
&&+D_p\phi_0D^p\phi_0+(K^m+6\beta\phi_0(\nu_m\Lambda\eps))^2+[\phi_0,\phi_A]^2\,.
\ee
All terms in (\ref{6dloc3}) are positive definite, except perhaps for the last two terms in the second and third lines.  However,  it is straightforward to show that these terms will be positive definite for all $\theta$ only if we choose $a=2/3$.  With this choice, (\ref{6dloc3}) becomes
\be\label{6dloc4}
&&\delta_\eps\Psi\, \overline{\delta_\eps\Psi}=\nn\\
&&\quad\left(D_\mu\phi_7-\sfrac{1}{3}F_{\lambda\rho}(\eps{\Gamma_\mu}^{7\lambda\rho0}\eps)\right)^2+\sfrac49\,v^\lambda v^\sigma F_{\mu\lambda}{F^\mu}_\sigma\nn\\
&&\quad+\cos^2\sfrac12\theta\left(\left(\phi_7-\sfrac{1}{6}f^{\bf N}\right)^2 +\sfrac29\cos^2\sfrac12\theta (f^{\bf N})^2+\left(1-\sfrac89\sin^2\sfrac12\theta\right) F^{{\bf N}+}_{\mu\nu}F^{{\bf N}+\mu\nu}\right)\nn\\
&&\quad+\sin^2\sfrac12\theta\left(\left(\phi_7-\sfrac{1}{6}f^{\bf S}\right)^2 +\sfrac29\sin^2\sfrac12\theta (f^{\bf S})^2+\left(1-\sfrac89\cos^2\sfrac12\theta\right) F^{{\bf S}+}_{\mu\nu}F^{{\bf S}+\mu\nu}\right)\nn\\
&&\quad+D_{\hat p}\phi_AD^{\hat p}\phi^A+(v^p D_{ p}\phi_A+\beta\phi^B\eps_{AB})^2+35\beta^2\phi_A\phi^A\nn\\
&&\quad+D_p\phi_0D^p\phi_0+(K^m+6\beta\phi_0(\nu_m\Lambda\eps))^2+[\phi_0,\phi_A]^2\,.
\ee

Examining (\ref{6dloc4}), we find that $\phi_A=0$ everywhere.  We also see that $f^{\bf N}=F^{{\bf N}+}=0$  except  at the south pole and $f^{\bf S}=F^{{\bf S}+}=0$  except at the north pole, which forces $\phi_7=0$ everywhere.   We also have that $v^\mu F_{\mu\nu}=0$, which along with the previous conditions are  enough to ensure that $F_{\lambda\rho}(\eps{\Gamma_\mu}^{7\lambda\rho0}\eps)=0$.
Finally, we find that $\phi_0$ is constant and $K^m=-6\beta\phi_0(\nu^m\Lambda\eps)$.  

These conditions allow for point-like anti-instantons on the north pole, defined with respect to $\omega^{\bf N}$, and point-like anti-instantons on the south pole, defined with respect to $\omega^{\bf S}$. 
Between the poles the situation is quite interesting.
   Here we would need $F=F^{{\bf N}-}=F^{{\bf S}-}$.  This then leads to the equations
\be\label{Feqs}
 F=-*(F\wedge\tilde\omega)\,,\qquad 0=F\wedge\bar\omega\,,
\ee
where we define $\bar\omega\equiv\sfrac12(\omega^N-\omega^S)$.  One can show that
\be\label{barom}
\iota_v\,\bar\omega=\iota_\theta\,\bar\omega=0
\ee
and that $\bar\omega$ spans the four-dimensional horizontal space defined by (\ref{barom}).  One can also show that
\be\label{dtom}
d\tilde\omega=3\sin\theta d\theta\wedge\bar\omega\,.
\ee
Hence, using the Bianchi identity, the first equation in (\ref{Feqs}) leads to
\be
d_A*F=-3\sin\theta F\wedge d\theta\wedge\bar\omega=0\,,
\ee
where we applied the second equation in (\ref{Feqs}) to do the last step.
Thus $F$ satisfies the Yang-Mills equations.  It also follows from the second equation in (\ref{Feqs}) and (\ref{barom}) that $\iota_\theta F=0$.
Therefore, we find for $\theta\ne0,\pi$ that there can be extended anti-instantons with respect to $\tilde\omega$ that lie in the co-dimension 2 horizontal space orthogonal to $v^\mu$ and the $\theta$ direction. 
{However since there are no non-trivial 2-cycles we conjecture that these extended instantons do not contribute to the path integral.}

The Yang-Mills action is zero for the point-like anti-instantons at the poles \cite{Iqbal:2003ds}, {which can be argued by using a simple scaling argument.}
 One could also consider extra terms in the Euclidean action with the form
\be
i\,\alpha \int \Tr[F\wedge F]\wedge \omega+i\,\beta \int \Tr[F\wedge F\wedge F]\,,
\ee
where $\alpha$ and $\beta$ are constants.  The first term is zero for the point-like anti-instantons.  It is also zero for the extended instantons discussed above, since $\omega$ is odd about the equator.    However, the last-term is proportional to the third Chern class.  Since the anti-instantons are point-like, on each pole this is equivalent to the third Chern class on $S^6$.  Hence one finds
\be
i\,\beta \int \Tr[F\wedge F\wedge F]={24\pi^3}i\,\beta\,\mathbb{Z}\,.
\ee
We will discuss the contributions of the instantons further at the end of this section.

\subsection{Calculation of determinants}

 Now we consider the calculation of the determinants around the trivial solution $A=0$, $\phi_0 = {\rm constant}$ and all other fields  zero. The one-loop contribution around 
  this configuration gives us the full perturbative part for the partition function on $S^6$. 
  
  The present calculation parallels Pestun's original calculation  \cite{Pestun:2007rz}  on $S^4$  (for the full equavariant version see \cite{Hama:2012bg}).
   Thus to avoid unnecessary repetition we mainly point out the differences between $S^4$ and $S^6$. In our presentation we follow  the notations from  \cite{Pestun:2007rz}. 

   We have to extend our supersymmetry complex (\ref{susyosr}) by the gauge fixing complex. Once this is done we can think of the extended supersymmetry 
    as an equivariant differential acting on a supermanifold with even coordinates $(A_M, \bar{a}_0, b_0)$ and odd coordinates $(\Upsilon_m, c, \bar{c})$.  
     As explained in  \cite{Pestun:2007rz} the calculation of the determinants boils down to the calculation of the equivariant index (with respect to rotation generated by $v$)
      of the operator $D_{10}$. 
      
But before proceeding we must show that $D_{10}$ is transversally elliptic in order for the index theorem to be applicable.  This is accomplished in a way that parallels Pestun's analysis \cite{Pestun:2007rz}, although we give a somewhat simpler method.  Effectively, $D_{10}$ appears in the expression
\be\label{elliptic1}
\sfrac{1}{2}\sum_{m=1}^7\Upsilon^m\nu_m\Gamma^{M'N'0}\eps F_{M'N'}+\tilde c\nabla^\mu A_\mu-c\nabla^\mu\LL_vA_\mu\,,
\ee
where the primed induces exclude the $0$ component and the terms involving the ghost fields fix the gauge.   Since the ellipticity of an operator is determined by its symbol $\algS$, we only need consider the leading derivatives, where we replace all $\nabla^\mu$ with $p^\mu$.  Upon doing this (\ref{elliptic1}) reduces to 
\be\label{elliptic2}
\sum_{m=1}^7\Upsilon^mW_m^{\mu N'} p_\mu A_{N'}+\tilde c\,p^\mu A_\mu-c\,p^\mu p^\nu v_\nu A_\mu+c\,p^2 v^{M'}A_{M'}\,,
\ee
where $W_m^{\mu N'}=\nu_m\Gamma^{\mu N'0}\eps=\nu_m\Gamma^{\mu}\Gamma^{N'}\Gamma^0\eps$ with this last equality holding because of the properties in (\ref{psrel}).  Using that $v_MW_m^{\mu M}=v_{M'}W_m^{\mu M'}=0$, we can rewrite (\ref{elliptic2}) as
\be\label{elliptic3}
&&\sum_{m=1}^7\Upsilon^mW_m^{\mu N'} p_\mu \tilde A_{N'}+(\tilde c-c\,p^\nu v_\nu)(p^\mu \tilde A_\mu+v_\mu p^\mu v^{M'}A_{M'})+c\,p^2 v^{M'}A_{M'}\nn\\
=&&\sum_{m=1}^8\Upsilon^mW_m^{\mu N'} p_\mu \tilde A_{N'}+\Upsilon_8v_\mu p^\mu v^{M'}A_{M'}+c\,p^2 v^{M'}A_{M'}\nonumber\\
=&&\{c,\Upsilon_m\}\algS\{v^{N'}A_{N'},\tilde A_{M'}\}^T\,,
\ee
where we have defined 
\be
\tilde A_{M'}&\equiv&A_{M'}-v_{M'}v^{N'}A_{N'}\nn\\
\Upsilon_8&\equiv&\tilde c-c\,p^\nu v_\nu\nn\\
W_8^{\mu N'}&\equiv&\eps\Gamma^{\mu}\Gamma^{N'}\Gamma^0\epsilon=\delta^{\mu N'}\,.
\ee
  The symbol $\algS$ of $D_{10}$ has the form of a $9\times9$ matrix and  it is clear from (\ref{elliptic3}) that $\det(\algS)=p^2\det(T)$, where ${T_m}^{N'}=W_m^{\mu N'} p_\mu$.   We now consider the  product
\be\label{elliptic4}
(T^T T)^{M'N'}&=& W_m^{\mu M'} p_\mu W_m^{\nu N'} p_\nu\nn\\
&=&\frac{1}{2}(\eps\Gamma_N\eps)(\eps\Gamma^0\tilde\Gamma^{M'}\slashed{p}\tilde\Gamma^N\slashed{p}\tilde\Gamma^{N'}\Gamma^0\eps)\,,
\ee
where we used the last equation in (\ref{psrel}) to get the second line in (\ref{elliptic4}).  Using (\ref{veq}), $v^{M'}\tilde A_{M'}=0$ and  $ v^\mu v_\mu+v_7^2=1$, it is  straightforward to show that
\be
(T^T T)^{M'N'}&=&\frac{1}{2}\left (p^2+v^\mu v_\mu(p_\perp^2-p_L^2)+v_7^2 p^2\right)\tilde g^{MM'}\nn\\
&=&(p_\perp^2+(1-v^\mu v_\mu)p_L^2)\tilde g^{M'N'}\,,
\ee
where $\tilde g^{M'N'}=\delta^{M'N'}-v^{M'}v^{N'}$ and $p^\mu=p_\perp^\mu+p^\mu_L$ with $p_\perp^\mu$  the component  orthogonal to $v^\mu$.  Hence $T^T T$ is diagonal, and thus we have
\be
\det(T)=(\det(T^T T))^{1/2}=(p_\perp^2+(1-v^\mu v_\mu)p_L^2)^4\,.
\ee 
Hence, $D_{10}$ is not elliptic at the equator, but is transversally elliptic with respect to the $U(1)$ action along $v$.

For the index,      only the north and south poles contribute and the appropriate index for $S^6$ is given by the following expression\footnote{We are grateful to Vasily Pestun for pointing out a mistake in (\ref{6dindex}) in an earlier version of this paper.}
  \bea\label{6dindex}
  {\rm ind} (D_{10})= &&2 + \left [ - \frac{1- \lambda^3}{(1-\lambda)^3}\right]_+ + \left [ - \frac{1- \lambda^3}{(1-\lambda)^3}\right]_- =\nonumber \\
   && =2 - (1-\lambda^3)(1+ 3\lambda + 6\lambda^2 + ....)
  - (1-\lambda^{-3})(1+ 3\lambda^{-1} + 6\lambda^{-2} + ....)=\nonumber  \\
  &&= - \sum\limits_{t=-\infty}^{\infty} 3 |t| \lambda^t ~. 
  \eea
  It means that one-loop contribution is given by the following infinite product
  \bea
  \prod_\alpha   \prod\limits_{t=1}^\infty (t +  i \langle \sigma, \beta\rangle )^{3t}~,
  \eea
 where $\alpha$ are the roots and the radius of $S^6$ has been absorbed into the dimensionless combination $\sigma=r\phi_0$ after Wick rotation.  Moreover it is a straightforward exercise to 
  write the fully equivariant answer when we take into account the full $T^3$ action on $S^6$ (see  \cite{Hama:2012bg} for the analogous calculation on $S^4$). 
  Thus the full equivariant perturbative result is given by the following matrix integral
\bea\label{6Dpert}
Z_{6D}^{pert} = 
 \int\limits_{\mathbf{t}} d\sigma~ e^{- \frac{16\pi^3 r^2\rho}{g_{YM}^2} {\rm Tr}(\sigma^2)} \prod\limits_{\alpha} \Upsilon (i \langle \sigma, \alpha \rangle;  \omega_1, \omega_2, \omega_3)~, 
\eea
 where we have introduced the following function 
\bea
\Upsilon_3 (x; \omega_1, \omega_2, \omega_3) = \frac{\prod\limits_{i,j,k=0}^\infty (i \omega_1 + j \omega_2 + k \omega_3 + x)}{\prod\limits_{i,j,k=1}^\infty (i \omega_1 + j \omega_2 + k \omega_3 - x)}
\eea
 with the  $\omega_i$ corresponding to the $T^3$ action on $S^6$ (in other words, they are squashing parameters for $S^6$). The case $\omega_1=\omega_2=\omega_3=1$ corresponds to 
  the round sphere.   We use the notation $\Upsilon_3 (x) = \Upsilon_3 (x; 1, 1, 1)$, where we find
\bea
 \Upsilon_3 (x) = x \prod\limits_{t=1}^\infty \frac{(t+x)^{\frac{1}{2}t^2 + \frac{3}{2} t +1}}{(t-x)^{\frac{1}{2} t^2 - \frac{3}{2} t +1}}
\eea
and thus we have
\bea\label{ups3rel}
  \prod_\beta \Upsilon_3 (i \langle \sigma, \alpha \rangle ) = \prod_\alpha  i \langle \sigma , \alpha \rangle \prod\limits_{t=1}^\infty (t +  i \langle \sigma, \alpha \rangle )^{3t}~.
\eea
 The term $\langle \sigma , \alpha \rangle$ comes from the Vandermonde determinant when we reduce the integration from the whole algebra to its Cartan subalgebra $\mathbf{t}$.

\subsection{{Conjecture} 
for the full answer}

At this point it would be natural to conjecture the full answer. In analogy with four dimensions  we would think that the full partition function on $S^6$ 
 is given by gluing two partition functions over $\mathbb{R}^6$ which count the point-like instantons (the solutions of the hermitian 
  Yang-Mills-Higgs system) whose  
  configurations are labeled by the 3D partitions.  In order for this conjecture to be valid we would
   need to show that $F=0$ everywhere except the poles of $S^6$. However in the previous section we showed that the localization locus allows for extended instantons  which have a nontrivial contribution to the action.  
    In six dimensions extended instantons can wrap non-trivial two cycles, as was 
  used in    \cite{Iqbal:2003ds} to establish the equivalence between a twisted $U(1)$ gauge theory on a toric Calabi-Yau and the topological vertex \cite{Aganagic:2003db}.
However on $S^6$ there are no non-trivial two-cycles, 
 so the extended instantons that we find are not precisely of this type.  Instead, these are instantons that are complementary to the point-like instantons, as they are well-defined everywhere except the poles where their defining conditions    break down.  From this point of view the instantons are extended along a cylinder and hence can be non-trivial.  In the future it would be interesting to find a nice expression for their contribution to the partition function, or alternatively, find some subtlety that rules out their existence.

\section{Matrix models}
\label{s-matrix}

We next discuss the large-$N$ behavior of the matrix models derived in the previous sections. 
 In the 't Hooft large-$N$ limit the instanton contributions are exponentially 
  suppressed. Thus we have to concentrate only on the perturbative part of the partition function  
   on the round spheres. 

\subsection{7D matrix model}\label{7Dmm}

The perturbative part of the partition function on the  round $S^7$  is given in (\ref{7Dpertpf}) which we can rewrite as the matrix integral
 of the infinite product  
 \be
 Z_{7D}^{pert}= \int\limits_{\mathbf{t}} d\sigma~ e^{- \frac{8\pi^4 r^3}{g_{YM}^2} {\rm Tr}(\sigma^2)}  
\prod_\alpha \left [ i \langle \sigma, \alpha \rangle \prod\limits_{t=1}^\infty \frac{(t+i
     \langle \sigma, \alpha \rangle)^{\frac{1}{6}t^3 + t^2 + \frac{11}{6} t + 1}}{(t-i
     \langle \sigma, \alpha \rangle)^{\frac{1}{6}t^3 - t^2 + \frac{11}{6} t - 1}} \right ]~,
 \ee
 where $\alpha$ are the roots. Thus the one-loop contribution is given in terms of the following infinite product
\bea
{\cal P}(x)= x \prod\limits_{t=1}^\infty \frac{(t+x)^{\frac{1}{6}t^3 + t^2 + \frac{11}{6} t + 1}}{(t-x)^{\frac{1}{6}t^3 - t^2 + \frac{11}{6} t - 1}}~.
 \eea
 This product is divergent where the divergent piece is give by the following expression
 \be
  \log {\cal P}(x) = \sum\limits_{t=1}^\infty \left ( \frac{1}{3} x^3 - x^2 + \frac{t^2}{3} x + \frac{11}{3} x \right ) + {\rm convergent~~ part}~.\label{7d-divp}
 \ee
  Thus the regularized version of our infinite product is given by the Weierstrass representation of the quadruple sine \cite{koyama2005}
 \be
 S_4 (x) = 2\pi x e^{-\zeta'(-2)} e^{-\frac{x^3}{18} + \frac{x^2}{2} - \frac{11}{6} x}  \prod\limits_{t=1}^\infty \left [
 \frac{(1+\frac{x}{t})^{\frac{1}{6}t^3 + t^2 + \frac{11}{6} t + 1}}{(1-\frac{x}{t})^{\frac{1}{6}t^3 - t^2 + \frac{11}{6} t - 1}}
  e^{-\frac{x^3}{9} + x^2 - \frac{t^2}{3} x - \frac{11}{3} x} \right ]~.
 \ee
  One  way to think about this regularization is that it introduces a cut-off in the mode expansion of 
   of the divergent part, stopping  the  expansion  at $n_0 = \pi^2 \Lambda^3 r^3$.  Thus summing over the roots we have
   \be
    \sum\limits_{\alpha}  \log\, {\cal P} (i
     \langle \sigma, \alpha \rangle) = - \pi^2 \Lambda^3 r^3 \sum\limits_{\alpha} (i
     \langle \sigma, \alpha \rangle)^2 + {\rm convergent~~ part}~,
   \ee
    where the cubic and linear terms in (\ref{7d-divp}) are equal to zero due to the 
  Weyl reflection symmetry.
     Using the identity $\sum\limits_{\alpha} (
     \langle \sigma, \alpha \rangle)^2 = 2 C_2({\rm adj}) {\rm Tr} (\sigma^2)$ we get
 \be
    \sum\limits_{\alpha}  \log\, {\cal P} (i
     \langle \sigma, \alpha \rangle) = 2  \pi^2 \Lambda^3 r^3 C_2({\rm adj}) {\rm Tr} (\sigma^2)  + {\rm convergent~~ part}~,
     \ee
 where $C_2({\rm adj})$ is the second Casimir of the adjoint representation ( $C_2$ is normalized to $1/2$ 
  for the fundamental representation).  The divergent part of the infinite product is proportional to ${\rm Tr} (\sigma^2)$
  and thus can be absorbed  into the redefinition of the coupling constant as 
  \be
   \frac{1}{g_{\rm eff}^2} = \frac{1}{g_{0}^2} - \frac{\Lambda^3}{4\pi^2} C_2 ({\rm adj})~.
  \ee
 where $g_0$ is the bare Yang-Mills coupling.
 From now on we  assume that all products are regularized and use $g_{YM}$ to mean the effective Yang-Mills coupling.

 Under the product over the roots the quadruple sine collapses to a function involving only  triple sines and standard sines, as in (\ref{S4-S3}). 
 Furthermore using the properties of the triple sine (e.g., see \cite{Minahan:2013jwa} for a summary) we 
 can rewrite the matrix model as follows
 \be
 Z_{7D}^{pert}= \int\limits_{\mathbf{t}} d\sigma~ e^{- \frac{8\pi^4 r^3}{g_{YM}^2} {\rm Tr}(\sigma^2)}  
\prod_\alpha \sinh ( \pi
     \langle \sigma, \alpha \rangle ) e^{ f(i
     \langle \sigma, \alpha \rangle)}~,\label{7Dmatmodel1}
 \ee
  where the function $f(x)$ introduced in \cite{Kallen:2012cs} is defined as follows
  \bea
 f(x) = \frac{i \pi x^3}{3} + x^2 \log (1- e^{-2\pi i x}) + \frac{ix}{\pi} \Li_2 (e^{-2\pi i x}) + \frac{1}{2\pi^2} \Li_3 (e^{-2\pi i x}) - \frac{\zeta(3)}{2\pi^2}
\eea
 with the property 
 \bea
  \frac{d f(x)}{d x} = \pi x^2  \cot (\pi x)~.
 \eea
 By looking  at the large $\sigma$ asymptotics 
   of the integrand (see \cite{Minahan:2013jwa})  we see that  the integral is convergent  and that (\ref{7Dmatmodel1}) is well-defined. 

 We next discuss the large $N$-behaviour of our matrix model. 
 Introducing the seven-dimensional 't Hooft coupling as
  \bea
 \lambda = \frac{ g_{YM}^2 N}{r^3}
 \eea
 and specializing to $SU(N)$ we can rewrite the matrix model (\ref{7Dmatmodel1}) in terms of the eigenvalues $\phi_i$
\bea
 \int \prod\limits_{i=1}^N d\phi_i \exp \Big (  - \frac{8 \pi^4 N}{\lambda} \sum\limits_i \phi_i^2 + \sum\limits_{i \neq j} \sum\limits_i \log (\sinh (\pi (\phi_i - \phi_j))) + f (i (\phi_i - \phi_j)) \Big) ~.
\eea

In large $N$ limit the partition function is dominated by the saddle point
\bea
 \frac{16 \pi^4 N}{\lambda} \phi_i = 2\pi \sum\limits_{i\neq j} (1 - (\phi_i - \phi_j)^2) \coth (\pi (\phi_i - \phi_j))
\eea
 This matrix model is very similar to the 5D matrix model for a pure  vector multiplet  and has qualitatively  the same behavior. In  \cite{Kallen:2012zn}  the matrix model for the 5D vector multiplet with any number of hypermultiplets 
  in the fundamental representation was considered. It is a straightforward exercise to generalize that analysis to the present matrix model. There it was argued  that the eigenvalues are spread over a finite extent   as $\lambda\to\infty$ due to a long distance attraction between the eigenvalues.  Therefore, all
    $\phi_i$ and $\phi_i - \phi_j$ are finite in this limit 
   and thus the free energy scales as $N^2$.

\subsection{6D matrix model}

 Using (\ref{6Dpert}) and (\ref{ups3rel}) we can write the perturbative part of the partition function for 6D super Yang-Mills on the round $S^6$ as  the matrix integral of the infinite product  
\bea
Z_{6D}^{pert} = 
 \int\limits_{\mathbf{t}} d\sigma~ e^{- \frac{16\pi^3 r^2 }{g_{YM}^2} {\rm Tr}(\sigma^2)}  \prod_\alpha  \left [ i \langle \sigma , \alpha \rangle
  \prod\limits_{t=1}^\infty  (t+i \langle \sigma , \alpha \rangle)^{3t} 
 \right ]~,
\eea
where $\alpha$ stands for the roots. The log of the  infinite product is logarithmically  divergent, and thus requires regularization.  
 As before we can regularize the product by introducing a cut-off for the modes $n_0 =\Lambda r$ in the divergent part and absorb this divergent part by an appropriate 
  redefinition of the coupling constant  
\bea\label{6Dgeff}
 \frac{1}{g_{eff}^2} = \frac{1}{g_{0}^2} - \frac{3 C_2 ({\rm adj})}{8 \pi^3 r^2} \log (\Lambda r e^{\gamma})~,
\eea
 where  $g_0$ is the bare coupling and  $\gamma$ is the Euler constant. From now on we assume that the product is regularized and the divergence is absorbed into the Yang-Mills coupling.  
Following \cite{Pestun:2007rz} and \cite{Russo:2012ay} we introduce the function  $H(x)$ as   
\bea
 H(x) =   \prod\limits_{n=1}^\infty \Big ( 1 + \frac{x^2}{n^2} \Big )   e^{-\frac{x^2}{n}}~. 
\eea
  Thus the 6D matrix model for round sphere can be written as 
 \bea
 Z_{6D}^{pert} = 
 \int\limits_{\mathbf{t}} d\sigma~ e^{- \frac{16\pi^3 r^2}{g_{YM}^2} {\rm Tr}(\sigma^2)} \prod\limits_{\alpha} i \langle \sigma, \alpha \rangle
 H^{3/2} ( \langle \sigma, \alpha \rangle)~,
 \eea
  This matrix model is similar to the matrix model for a pure  $N=2$ vector multiplet on $S^4$, the only difference being the power of the $H(x)$.
  The Barnes function $H(x)$ has  the following asymptotics   
 at large real $x$
  \bea
   \log H (x) \sim  - x^2 \log |x| e^{\gamma - \frac{1}{2}} + O(\log x)~.
  \eea
  Therefore we can conclude that matrix integral converges and the matrix model is well-defined. 
       
       Next we look at the large $N$-limit. Introducing the 6D 't Hooft coupling
  \bea
 \lambda = \frac{ g_{YM}^2 N}{r^2}
 \eea
  and  assuming the gauge group to be $SU(N)$ we can rewrite the matrix model in terms of eigenvalues 
   \bea\label{6Dmm}
 Z_{6D}^{pert} = 
 \int\prod\limits_{i=1}^N d\phi_i e^{- \frac{16\pi^3 N}{\lambda} \sum\limits_i \phi_i^2 } \prod\limits_{i \neq j} \prod\limits_i  (\phi_i- \phi_j)
  H^{3/2} (\phi_i - \phi_j) ~.
 \eea
  This can also be solved by saddle point.  
  Exploiting the similarity with the pure 4D $\NN=2$ theory we can borrow the results from \cite{Russo:2012ay} to study the behavior of (\ref{6Dmm}).     One clear feature is that at large $N$ the free energy behaves as $N^2$ for large $\lambda$.  
  
  It is also possible in the 6D theory, as in the pure 4D $\NN=2$ theory,  to have an effective square coupling which is negative.  One can see this in  the 4D case  by starting with a massless adjoint hypermultiplet, which enhances the supersymmetry to $\NN=4$.  Turning on the mass, it acts as a UV cutoff and runs the square inverse of the effective  coupling downward, until it eventually crosses zero.    But the corresponding matrix model is still well-defined.  In the 6D case we do not have the luxury of a field mass to serve as a cutoff.  Nevertheless, we can still push $g_{eff}^{-2}$ into negative territory in  (\ref{6Dgeff}) by choosing $\Lambda$ large enough, and still have a well-defined matrix model. Recently, it was argued that a similar phenomenon happens for the pure 5D $\NN=1$ theory where the effective squared coupling can be negative \cite{anton}.  One can then argue the same for the 7D theory.
 
\section{Summary}
\label{s-summary}

In this work we have constructed maximally supersymmetric theories on $S^d$ with $ d\leq 7$ and described their localization.
In the case of  $S^6$ and $S^7$ 
we have derived the complete localization locus and calculated the full perturbative answer.  The perturbative results are described in terms of matrix models for which we have briefly
  discussed  their large $N$-behaviour. 
 Unfortunately, the localization locus  for both $S^6$ and $S^7$  appears difficult to analyze
  explicitly. In particular it seems that  extended instanton solutions are not ruled out. However using the analogy with previous localization results we conjecture that for $S^6$
   only solutions which sit on fixed points of $U(1)$ actions contribute to the path integral and there is no non-trivial topology to support the extended instantons. For $S^7$
    we conjecture that the instantons sit on closed orbits of the $U(1)$ action and again there is no non-trivial topology to support the extended instantons. These issues definitely 
     require further detailed study.

{There are many directions along which our results can be generalized  and extended. For example the calculation on $S^7$ can be easily extended to any toric 7D Sasaki-Einstein 
manifold in the same fashion as in the 5D case \cite{Qiu:2014oqa}. It would also be interesting to study the different versions of matrix models which can arise from 
 the inclusion of Chern-Simons observables in 7D.}

An interesting application of the six-dimensional theory is as a laboratory for studying little string theory \cite{Seiberg:1997zk,Aharony:1998ub} (see also  \cite{Aharony:1999ks,Kutasov:2001uf} for  reviews).  Little string theory is believed to be a valid UV completion of  six-dimensional super Yang-Mills. The strings are essentially the instantons which have co-dimension two in six dimensions.  The relation between the Yang-Mills coupling and the string mass is
\be
\frac{1}{g_{YM}^2}=M_s^2\,,
\ee
hence six-dimensional super Yang-Mills is trustworthy when $E g_{YM}\ll1$, where $E$ is the characteristic energy scale.  For the partition function on $S^6$, the only scale  that appears is $r$, hence the corresponding condition is that $g_{YM}/r\ll1$ and  one would normally think that the  instantons are suppressed.  This would be the case for the extended instantons, assuming that they contributed to the partition function.  But the point-like instantons are not suppressed in this limit and could contribute significantly to the partition function with $\theta$ dependent terms.  One interesting question is whether they could affect the $N^2$ dependence of the perturbative free-energy.   Another question is whether the sum over the point-like instantons  could be interpreted as a sum over ``short" little-strings.

The case of super Yang-Mills on $S^7$ is even more mysterious, as gauge theories in seven dimensions have no known   UV completion that does not contain gravity.  Here, the extended instantons are co-dimension three so the UV completion contains membranes.  In \cite{ArkaniHamed:2001ie} it was argued 
that the ``little m-theory" \cite{Losev:1997hx} does not de-couple from the bulk M-theory, so perhaps the UV completion is M-theory itself.  It would be nice if one could address this question further using the localized 7D Yang-Mills theory.

\section{Acknowledgements}
 We are grateful to Chris Hull, Anton Nedelin, Vasily Pestun, Jian Qiu and  Leonardo Rastelli for discussions. 
The research of J.A.M. is supported in part by
Vetenskapsr{\aa}det under grant \#2012-3269.
J.A.M. thanks the CTP at MIT  for kind
hospitality during the course of this work.
The research of M.Z. is supported in part by
Vetenskapsr{\aa}det under grant \#2014-5517, by the STINT grant and by the grant ``Geometry and Physics" 
 from the Knut and Alice Wallenberg foundation. 

\appendix\section{Conventions and useful properties}\label{conv}

\subsection{General dimensions}
We use 10-dimensional Majorana-Weyl spinors $\eps_\al$, $\Psi_\al$ {\it etc.}.  Spinors in the other representation are written with a tilde, $\tilde\eps^\al$, {\it etc.}  The 10-dimensional $\Gamma$-matrices are chosen to be real and symmetric, 
\be
\Gamma^{M\al\beta}=\Gamma^{M\beta\al}\qquad \tilde\Gamma^M_{\al\beta}=\tilde\Gamma^M_{\beta\al}\,.
\ee
Products of $\Gamma$-matrices are given by
\be
&&\Gamma^{MN}\equiv \tilde\Gamma^{[M}\Gamma^{N]}\qquad \tilde\Gamma^{MN}\equiv \Gamma^{[M}\tilde\Gamma^{N]}\nn\\
&&\Gamma^{MNP}\equiv\Gamma^{[M}\tilde\Gamma^{N}\Gamma^{P]}\qquad\tilde\Gamma^{MNP}\equiv\tilde\Gamma^{[M}\Gamma^{N}\tilde\Gamma^{P]}\,,\ \mbox{\it etc.}
\ee
We also have that $\Gamma^{MNP\al\beta}=-\Gamma^{MNP\beta\al}$, hence
\be
\eps\Gamma^{MNP}\eps=0\,
\ee
for any bosonic spinor $\eps$.

A very useful relation is the triality condition,
\be\label{triality}
\Gamma^M_{\al\beta}\Gamma_{M\gamma\delta}+\Gamma^M_{\beta\delta}\Gamma_{M\gamma\al}
+\Gamma^M_{\delta\al}\Gamma_{M\gamma\beta}=0\,.
\ee
We can use this to show that
\be
\eps\Gamma^M\eps\,\eps \Gamma_M\chi=0\,,
\ee
where $\chi$ is any spinor.  It immediately follows that $v^Mv_M=0$, where $v^M$ is the vector field
\be
v^M\equiv \eps\Gamma^M\eps\,.
\ee

We also use a set of pure-spinors, $\nu_m$ which satisfy the properties
\be\label{psrel}
\nu_m\Gamma^M\eps&=0&\nn\\
\nu_m\Gamma^M\nu_n&=&\delta_{mn}v^m\nn\\
\nu^m_\al\nu^m_\beta+\eps_\al\eps_\beta&=&\frac12 v^M\tilde\Gamma_{M\al\beta}\,.
\ee
They are invariant under an internal $SO(7)$ symmetry, which can be enlarged to $SO(8)$ by including $\eps$.

Defining
\be
\omega^{\mu\nu}\equiv (\eps\,\tilde\Gamma^{\mu\nu}\Lambda\,\eps)\,,
\ee
where $\Lambda\equiv\Gamma^{890}$ and using the Killing spinor equation in (\ref{KS}), one can show
\be
\nabla_{[\mu,}\omega_{\nu\lambda]}=0\,,
\ee
while
\be\label{divom}
\nabla^\mu\omega_{\mu\nu}=\frac{d-1}{r}\,v_{\nu}\,.
\ee

\subsection{Some useful relations in odd dimensions}

In odd dimensions we can set $v_\mu v^\mu=1$.  We have using (\ref{psrel}),
\be
v^\mu\omega_{\mu\nu}=0\,.
\ee
We define a one-form $\kappa_\mu$, such that $v^\mu\kappa_\mu=1$.  From (\ref{divom}) we see that 
\be
\kappa_\nu=\frac{r}{d-1}\,\nabla^\mu\omega_{\mu\nu}\,.
\ee
We also have the useful identity
\be\label{iden1}
(\eps\Gamma^{\mu\nu\lambda A0}\eps)(\eps{\Gamma_\mu}^{\sigma\rho B0}\eps)&=&
(\hat g^{\nu\sigma}\hat g^{\lambda\rho}-\hat g^{\nu\rho}\hat g^{\lambda\sigma}-\omega^{\nu\lambda}\omega^{\sigma\rho}-(\eps\Gamma^{\nu\lambda\sigma\rho0}\eps))\delta^{AB}
\nn\\
&&\qquad+[\hat g^{\nu\sigma}\omega^{\lambda\rho}-\hat g^{\nu\rho}\omega^{\lambda\sigma}-\hat g^{\lambda\sigma}\omega^{\nu\rho}+\hat g^{\lambda\rho}\omega^{\nu\sigma}]\ve^{AB}\,,
\ee
where $\hat g^{\mu\nu}\equiv g^{\mu\nu}-v^\mu v^\nu$.
From this it  follows that
\be\label{iden2}
(\eps\Gamma^{\mu\nu\lambda A0}\eps)(\eps{\Gamma_{\mu\nu}}^{\sigma B0}\eps)&=&
(d-3)(\hat g^{\lambda\sigma}\delta^{AB}+\omega^{\lambda\sigma}\ve^{AB})\,.
\ee
Another relation is
\be\label{iden3}
\nabla_\mu(\eps\Gamma^{\mu\nu\lambda A0}\eps)=0\,.
\ee
which follows from (\ref{KS}) and (\ref{psrel}).

\subsection{Other useful relations}

In six dimensions we have the relation
\be\label{6dsq}
&&(\eps \Gamma^{\lambda7\mu\nu0}\eps)(\eps {\Gamma_{\lambda}}^{7\mu\nu0}\eps)=(\eps \Gamma^{M7\mu\nu0}\eps)((\eps {\Gamma_{M}}^{7\mu\nu0}\eps)=\nn\\
&&\qquad \sin^2\theta\left(g^{\mu\sigma}g^{\nu\kappa}-g^{\mu\kappa}g^{\nu\sigma}+\sfrac14\varepsilon^{\mu\nu\sigma\kappa\lambda\rho}(\omega^{\bf N}_{\lambda\rho}+\omega^{\bf S}_{\lambda\rho})\right)\nn\\
&&\qquad-g^{\mu\sigma}v^{\nu}v^{\kappa}-v^{\mu}v^{\sigma}g^{\nu\kappa}+g^{\mu\kappa}v^{\nu}v^{\sigma}+v^{\mu}v^{\kappa}g^{\nu\sigma}
\ee
where $v^\mu=\eps\Gamma^\mu\eps$ and $\omega^{\bf N}_{\lambda\rho}$ and $\omega^{\bf S}_{\lambda\rho}$ are defined below (\ref{omrel}).

\bibliographystyle{JHEP}
\bibliography{refs}  
\providecommand{\href}[2]{#2}\begingroup\raggedright\endgroup

\end{document}